%% file: main.tex
\documentclass{article}
\usepackage[preprint]{neurips_2020}
\usepackage{geometry}
\usepackage{graphicx}
\usepackage{qtree}
\usepackage{booktabs}
\usepackage{siunitx}
\usepackage{ragged2e}
\usepackage{rotating}
\usepackage{adjustbox}
\usepackage{caption}
\usepackage[T1]{fontenc}
\usepackage{pdflscape}
\usepackage{nameref}
\usepackage{pdfpages}
\usepackage{float}
\newcolumntype{R}[1]{>{\RaggedRight}p{#1}}
\usepackage{amssymb}
\usepackage{xcolor}
\usepackage{verbatim}
\usepackage{subcaption}
\usepackage{array}
\usepackage{natbib}
\usepackage{hyperref}
\usepackage{mathtools}
\usepackage{cleveref}
\usepackage{svg}
\usepackage{transparent}
\usepackage[toc,title,page]{appendix}
\definecolor{orangefull}{RGB}{230, 159, 0}
\colorlet{orange}{orangefull!20!white}
\definecolor{bluefull}{RGB}{86, 180, 233}
\colorlet{blue}{bluefull!20!white}
\definecolor{green}{RGB}{0, 158, 115}
\colorlet{green}{green!20!white}
\title{Mind the (spectral) gap: How the temporal resolution of wind data affects multi-decadal wind power forecasts}
\author{%
  Nina Effenberger\thanks{corresponding author} \\
  Cluster of Excellence Machine Learning\\
  University of Tübingen\\
  72076 Tübingen \\
  \texttt{nina.effenberger@uni-tuebingen.de} \\
  \And
  Nicole Ludwig \\
  Cluster of Excellence Machine Learning\\
  University of Tübingen\\
  72076 Tübingen \\
  \texttt{nicole.ludwig@uni-tuebingen.de}
  \And
  Rachel H. White \\
  Department of Earth, Ocean and Atmospheric Sciences\\
  University of British Columbia\\
  Vancouver, BC Canada V6T 1Z4 \\
  \texttt{rwhite@eoas.ubc.ca}
}
\begin{document}

\maketitle

\begin{abstract}
To forecast wind power generation in the scale of years to decades, outputs from climate models are often used. However, one major limitation of the data projected by these models is their coarse temporal resolution - usually not finer than three hours and sometimes as coarse as one month. Due to the non-linear relationship between wind speed and wind power, and the long forecast horizon considered, small changes in wind speed can result in big changes in projected wind power generation. Our study indicates that the distribution of observed 10min wind speed data is relatively well preserved using three- or six-hourly instantaneous values. In contrast, daily or monthly values, as well as any averages, including three-hourly averages, are almost never capable of preserving the distribution of the underlying higher resolution data. Assuming that climate models behave in a similar manner to observations, our results indicate that output at three-hourly or six-hourly temporal resolution is high enough for multi-decadal wind power generation forecasting. In contrast, wind speed projections of lower temporal resolution, or averages over any time range, should be handled with care. 
\end{abstract}
%\ioptwocol

\section{Introduction}
\label{introduction}
%wind is an important source of power and difficult to forecast, accurate very long-term forecasts are important 
Wind is one of the main sources of renewable power and its utilisation is on the rise in many countries, \citep[e.g. ][]{soares2020current}. It is therefore of uttermost importance to have reliable wind power forecasts in the range of years to decades (i.e. a turbine's lifetime) for site assessment and reliable future power supply \citep{copernicus}. However, the high variability and stochasticity of weather and wind introduces uncertainty that makes multi-decadal planning difficult. Furthermore,  computational complexity limits the resolution of forecasts; the temporal and spatial
 resolution \footnote{Notice that temporal and spatial resolution are often directly linked e.g. \cite{courant1928partiellen}} of long-term and multi-decadal forecasts are therefore usually much coarser than that of short-term forecasts \citep{eyring2016overview}. 

%introduce research question whether wind variability plays a role for long-term wind power forecsating 
But do the low temporal resolution outputs from e.g. climate models provide wind speeds representative of the true site-specific high resolution wind speed distribution? An indicator that this could be the case is the so called wind power spectral gap \citep{van1957power}. 
%  wind power spectrum and spectral gap and turbulence
Wind speed variability can be assessed in the frequency domain in terms of power spectra by (Fourier-) decomposing high-resolution wind speed observations. This decomposition reveals an amplitude gap between high and low frequencies where the strong variability in high frequencies (on the order of seconds to minutes) is associated with turbulence, while strong low frequency variability (hours to days) is associated with synoptic weather systems \citep{van1957power},\citep{stull1988introduction}. In the gap between the synoptic weather and turbulence peaks in the frequency spectrum are frequencies with little variability; this was found to be in the range of $10^{-3}Hz (\sim 17min)$ to $10^{-4}Hz (\sim 3h) $ by \cite{van1957power}. However, other research suggests the gap is smaller \citep{kang2016spectral} or may not exist at all \citep{larsen2016full}. This spectral gap is described in many research papers \citep{horvath2012sub}, \citep{kang2015intra},  \citep{larsen2016full}, \citep{lopez2021effects} indicating that observations that fall into the corresponding frequencies do not add much knowledge. Given the inconsistency of the width of this spectral gap, the question remains of what temporal resolution we should aim for in multi-decadal wind power forecasting. 

%Small changes make a difference
Due to the non-linear relationship between wind speed and wind power, small changes in the wind speed distribution can lead to significant changes in wind power generation. However, if the underlying wind speed characteristics are preserved, lower-resolution data is preferred for multi-decadal forecasting to reduce required storage space and potentially also computational costs.
%Instead of forecasting wind speeds temporarily precisely our main goal is to model the wind speed distribution accurately, e.g. to estimate how much wind power a turbine can generate over its lifetime.

%A very short paragraph on average trends 
Additionally, multi-decadal wind speed projections should also account for climate change and interannual climate variability \citep{pryor2018interannual}. Several studies indicate that climate change will affect average wind speeds \citep{mcvicar2012global}, \citep{tobin2015assessing}) as well as wind speed variability \citep{pryor2010climate}, \citep{tobin2016climate}, \citep{dunn2019changes}, \citep{jeong2019projected}, \citep{ringkjob2020short} which will impact wind power generation. In general, it must be assumed that current wind conditions may not be representative of future wind conditions \citep{jung2019changing}; we thus often rely on output from climate models to help inform about future potential wind power. But which data (resolution) do we need for meaningful wind speed and power predictions? 

There seems to be agreement among researchers that certain temporal resolutions are \textit{too low}. It is therefore common to account for additional variability using so-called downscaling techniques \citep{pryor2010climate}. In the past, different statistical downscaling approaches have been introduced, e.g. \citep{von1999use}, \citep{tobin2015assessing}, \citep{shin2018novel}, as well as dynamical downscaling using regional climate models such as CORDEX \citep{giorgi2015regional}, used by e.g. \cite{davy2018climate} and \cite{yang2022climate}. 
%However, many climate datasets are coarsely resoluted and it is a common approach to simulate wind speed variability to account for sub-resolution variability\cite{pryor2010climate}. Approaches to adjust wind speed variability include inflation \cite{von1999use} where variability is increased by multiplication by a specified factor and randomization where introducing noise increases the variability. Other approaches include reproducing past wind speed variability statistically \cite{tobin2015assessing} or modelling the daily wind speed using a Weibull distribution \cite{shin2018novel}. 
%Weibull
The question remains, however: what temporal resolution of wind speed data is \textit{high enough}?
%\cite{khan2022bayesian} have shown that a three-parameter Weibull distribution is best suited for low wind speeds with high skewness. %Furthermore it has been observed that when fitting a Weibull distribution to observations, the parameters of the distribution can vary with season and site\cite{lun2000study}.

% concrete question, also introduce averaged vs instantaneous
The output from climate models are often available in various temporal aggregations, where some datasets represent temporal averages and other datasets consist of instantaneous values. In the CMIP6 datasets \citep{eyring2016overview}, one of the most widely used set of global circulation models (GCMs), wind projections are available as temporal means and instantaneous values \citep{cmip6}. Currently, however, little focus has been placed on the choice of data and many studies are performed without explicitly stating whether averages or instantaneous values are being used. 

With this study we show that the type of data (instantaneous vs averages) influences the wind speed distributions and thus the estimated wind power generation. We also conduct analysis to determine whether there is a temporal resolution that is \textit{high enough}, i.e. for which added temporal resolution provides little additional information and accuracy. We give recommendations regarding the choice of data and the temporal resolution that downscaling techniques, and climate model outputs, should aim for.
%importance of analyzing real data, i.e. measurements on hub height and lack thereof
%All of the reviewed studies give an insight into the role of wind speed variability for wind power forecasting and reveal that this variability might change due to climate change. However, it is still not clear how wind speed variability will change with climate change and to the best of our knowledge is has yet not been investigated which importance the coarse resolution plays for wind power generation. In order to fill this research gap we compute mean hub-height wind speeds on different temporal scales and compare their theoretical power output directly. 
To do this, we conduct empirical data analysis using data from eight different mid-latitude sites across Europe and North America. In \Cref{data} we describe the data and the methodology used. Our results are presented in \Cref{results} and we discuss their implications in \Cref{discussion}. Finally, we conclude in \Cref{conclusion}.

\section{Methods}
\begin{figure}    
\centering
\includegraphics[width=0.5\textwidth]{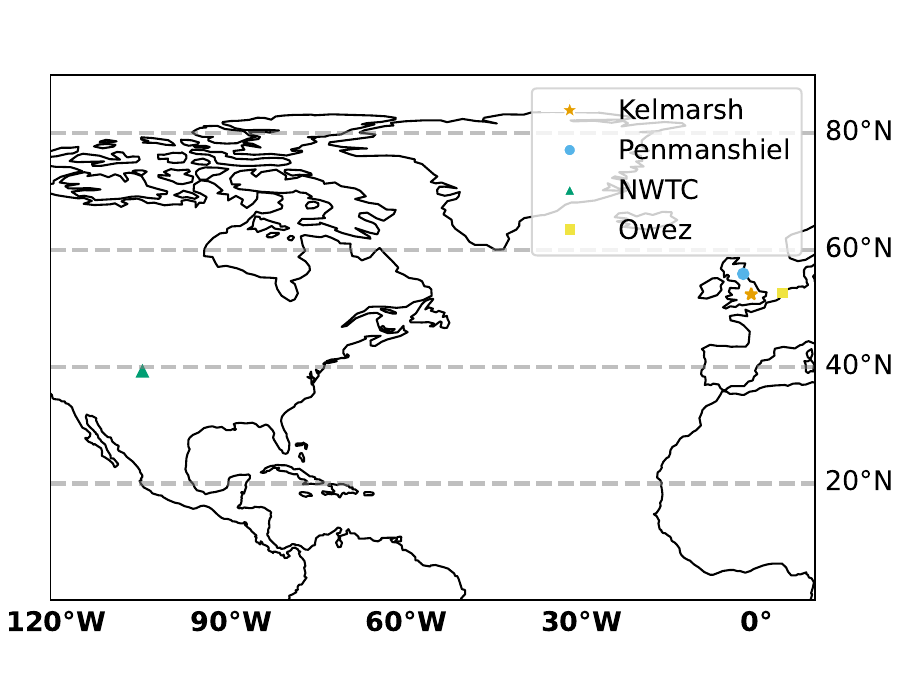}
        \caption{Locations of the two wind farms in the UK and the tall towers in central North America and the Netherlands. Our hub-height observation locations include one mountainous site (NWTC), one off-shore site (Owez), one coastal site (Penmanshiel) and one site on flat terrain (Kelmarsh). Penmanshiel and Kelmarsh are wind farm sites, their wind speeds are therefore influenced by wake effects of the surrounding turbines \citep[e.g. ][]{gonzalez2012wake} and we only used one turbine for our evaluations. }
    \label{fig:locations}
\end{figure}

%Goal: Find out whether averaging/excluding data results in different wind power generation 
To investigate how aggregating wind speed data affects the wind speed distribution we use both parametric and non-parametric approaches. In the main manuscript we focus on observations of turbine-hub-height winds from four sites, with locations shown in \Cref{fig:locations}. We use these as our primary data as they are hub-height data; however, all four sites have a relatively limited observational period of only a few years (see \Cref{tab:hub-heights}). We thus also use 10$m$ wind speeds from an additional four observational met masts from locations across Germany (see \Cref{appendix:locations-DWD} for locations) that have between $18$ to $34$ years of data available. These data show very similar results to the hub-height datasets (analysis presented in the supplementary material in \Cref{appendix}). We also use these longer datasets to analyze multi-decadal tendencies of wind power generation in \Cref{implications}. In the following we first give a description of the data used and then describe our methodology. 

\subsection*{Data}\label{data}
We investigate wind speed observations using open-source high met mast wind data of four mid-latitude locations. All of the wind speeds are either measured at wind turbine hub-height directly, i.e. by nacelle anemometers (sites Penmanshiel and Kelmarsh) or by high met masts (sites NWTC and Owez). The observation heights are between $59m$ and $116m$ and all of the measurements are provided as 10min averages, a very common aggregation-level of wind resource data \citep{harper2010guidelines}. In \Cref{tab:hub-heights} we present static information of the observation sites. Static information of the longer datasets (at 10$m$ height) is presented in \Cref{appendix:hub-heights-appendix}. Unless specified otherwise, the following abbreviations for the datasets can be found in the top left corner of figures: a) Kelmarsh, b) Penmanshiel, c) NWTC, d) Owez, e) Aachen, f) Zugspitze, g) Boltenhagen, h) Fichtelberg.

To bring the data to a format where we can compare different temporal resolutions, we pre-process the data by excluding all days where at least one observation is missing. We then average the 10 min wind speed observations to three-hourly, six-hourly and daily data: the $n$'th wind speed value $w_n$ in the time-series averaged over $t$ consecutive time steps is computed as
\begin{equation}
    w^t_n = \frac{1}{t} \sum_{i=nt}^{nt+t}w^{1}_i,
\end{equation}
%Given a time series of length $l$ this procedure results in averaged time series of length $\frac{l}{t}$ where our data pre-processing makes sure that this is a valid length. 
%Instantanous values
where $t=18$ (three-hourly), $36$ (six-hourly) and $144$ (daily) respectively; for $t=1$ we get the original 10 min resolution time series $w^{1}$.
In addition to calculating averages, we also consider wind speed time series of lower resolution, which we call instantaneous values. To do so, we use wind speed measurements every $t$'th time step only and exclude all other wind speed measurements $w_i$ where $i\neq nt$. The results are eight observations per day (every three hours), four observations per day (every six hours) and one observation per day respectively.

\begin{table}
\small
    \centering
    \begin{tabular}{
    |p{0.11\textwidth}|
    >{\centering\arraybackslash}p{0.12\textwidth}|>{\centering\arraybackslash}p{0.08\textwidth}|>{\centering\arraybackslash}p{0.15\textwidth}|>
    {\centering\arraybackslash}p{0.1\textwidth}|>
    {\centering\arraybackslash}p{0.09\textwidth}|>{\centering\arraybackslash}p{0.18\textwidth}|}
    \hline
         Name & Country\newline(Lat, Lon) & Temporal resolution in min& Observation period \newline(Number of days)& Observation height in $m$ & Mean wind speed (variance) in $\frac{m}{s}$ & Data source\\ \hline
         Kelmarsh& United Kingdom (52.40, -0.95) & 10 & 2016-2021 (1687) & 78.5 & 6.25 (7.70) &\cite{plumley_charlie_2022_5841834}\\ \hline
         Penmanshiel & United Kingdom (55.90, -2.31)& 10& 2016-2021 (1522) & 59 & 7.27 (15.30)& \cite{plumley_charlie_2022_5946808} \\ \hline
         Owez & Netherlands (52.61, 4.39)& 10 &2012-2017 (1148)& 116 & 4.82 (13.02)&\cite{ramon2020tall}, \cite{owez} \\ \hline
         NWTC &United States\newline (39.21, -105.23) & 10 &2005-2010 (1083)& 87& 8.37 (16.55)     &\cite{ramon2020tall}, \cite{NWTC} \\ \hline
    \end{tabular}
    \caption{Static data of the four different sites with hub-height measurements. Our chosen datasets cover a large range of observation heights as well as different mean wind speeds and variances.}
    \label{tab:hub-heights}
\end{table}

%Main idea: compare distributions of real wind data, averaged on different levels and instantenous at different time points

%While this is a very strict limitation it resulted in more than three years of data for three of the datasets and almost three years of data for the last dataset. For Kelmarsh, Penmanshiel, Owez and NWTC we evaluated 1687, 1522, 1148 and 1083 from originally 1825, 1627, 1918 and 1733 days of data respectively. 
%4. Kolmogorov-smirnov statistics on acumulated densities 
\subsection*{Comparing wind speed distributions}
To determine whether wind speed distributions from data of different temporal resolution are  statistically different, we compute pairwise Kolmogorov-Smirnov test statistics of cumulative density distributions 
\begin{equation}
    F_W(w) = P(W \leq w).
\end{equation}
%Notice that the original wind data, $w$, not the parameterized distributions, are used here. 
The Kolmogorov-Smirnov statistic $D$ is given by:
\begin{equation}
    D = \sup\limits_{w} |T(w)-S(w)|
\end{equation}
where $w$ are the wind speed values, $T$ and $S$ are the wind speed distributions to be compared, and the supremum, $\sup$, is the largest value of the set of values $|T(w)-S(w)|$ across all $w$. The Kolmogorov-Smirnov test only takes the largest absolute difference between the two distributions across all $w$ values into account and we identify a statistically significant difference if the $p$-value of an individual test is $p\leq 0.05$. 

\subsection*{Modeling wind speeds using Weibull distributions}
%0. Data 

%1. model density of wind speeds using weibull 
The Kolmogorov-Smirnov test can tell us whether wind distributions of different temporal resolution differ; in order to quantify the differences found, we model the wind speeds, $w$, as Weibull distributions. This is done by fitting the parameters of a three-parameter Weibull distribution,
\begin{equation}
    f(w; \beta, \lambda, \theta)= \frac{\beta}{\lambda}(\frac{w-\theta}{\lambda})^{\beta-1}e^{-(\frac{w-\theta}{\lambda})^\beta}
\label{eq:weibull}
\end{equation}
to the different temporal resolution datasets. The three-parameter Weibull distribution described in \Cref{eq:weibull} is defined by $w\ge\theta$ and $f(w; \beta, \lambda, \theta)=0$ for $w<0$ where $\beta>0$ is the shape parameter, $\lambda>0$ is the scale parameter and $\theta$ is the location parameter of the distribution which equals the lowest possible value of the distribution. For $\beta \approx 3$, the Weibull distribution approximates a Gaussian distribution, while $3>\beta \geq 1$ corresponds to a right-skewed distribution, and $\beta>3$ corresponds to a left-skewed distribution, for $\beta<1$ the density values are steadily decreasing with increasing $w$. $\lambda$ represents the variability, i.e. smaller values of $\lambda$ are associated with less variability \citep{rinne2008weibull}. We fit the parameters using Maximum Likelihood Estimation (MLE). This approach requires maximizing the likelihood function 
\begin{equation}
    L = \prod_{i=1}^n f(w_i; \beta, \lambda, \theta).
\end{equation}

We can then evaluate the change of the parameters when wind speeds are averaged or discarded to produce datasets with different temporal resolution. While Weibull distributions are commonly used to model wind speed distributions \citep[e.g. ][]{mert2015statistical}, we additionally use a generalized Gamma distribution to test the sensitivity of our results to the choice of underlying distribution. The conclusions are unchanged (figures shown in the appendix) and we therefore consider our results to be insensitive to the distribution choice. 

%2. qq plots to show that this is valid 
\subsection*{Validating the Weibull parametrization}
Using a kernel density estimation we confirm that Weibull distributions are a reasonable representation of our data, with the exception of monthly averages. The kernel density estimator $\hat{f}$ of an unknown density $f$ at a point $x$ is defined by 
\begin{equation}
    \hat{f}_h(x)=\frac{1}{nh}\sum_{i=1}^n K(\frac{x-x_i}{h}),
\end{equation}
where we choose $K$ to be the Gaussian kernel 
\begin{equation}
    K(x,h)=\exp(-\frac{x^2}{2h^2}).
\end{equation}
The band-width $h$ is selected using Scott's rule \citep{scott2015multivariate}. 

Figures \ref{appendix:all-kdes} and \ref{appendix:KDE-instantaneous} show kernel density estimations for averaged and instantaneous wind speeds respectively; it is clear that monthly values can not be described by a Weibull distribution and thus we do not include this temporal resolution in subsequent analysis.
To validate the fit of the Weibull distributions to the original wind speed distributions we generate quantile-quantile plots of the observations of length $\frac{l}{t}$ against $\frac{l}{t}$ randomly drawn samples from the corresponding Weibull distribution; these plots are shown in \Cref{appendix:QQ-plot-Weibull-avrg} and \Cref{appendix:QQ-plot-Weibull-inst}. The Weibull distributions generally provide a good fit to the data, although in some locations the fit is less good at higher wind speeds.

%3 Trend of parameters of the weibull distributions
%\subsection*{Compute trends of Weibull parameters}
%Given the fitted distributions of wind speeds, we can compute how the parameters of the Weibull distribution change when averaging or excluding data. To do so, we take the initial parameters $p^1$ given by the distribution fitted to 10min wind speeds and compute the fractional change of the parameters. The fractional change $c$ of averaging over every consecutive $t$ values is given by: 
%\begin{equation}
    %c^t = \frac{p^t}{p^1}.
%\end{equation}

\subsection*{Power generation and transferability of results}
\begin{figure}
    \centering
\includegraphics[width=0.5\textwidth]{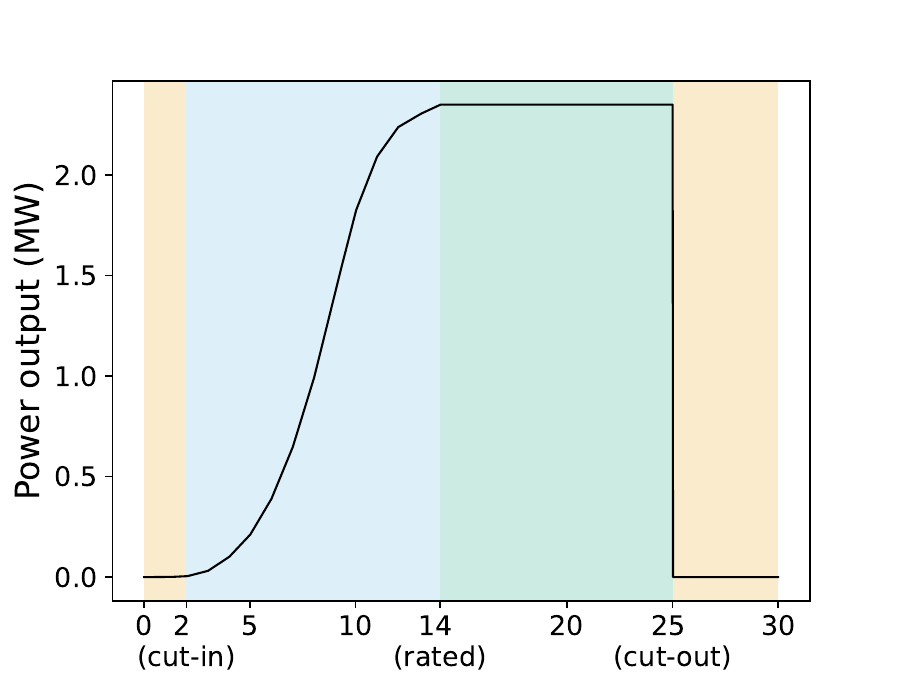}
    \caption{Wind power curve of Enercon E92/2350 turbine. The relationship between wind speed and wind power can be roughly divided into four different regions: No power is generated if the wind speed is below the cut-in wind speed or if the wind speed is above the cut-out wind speed where the turbine is shut down to protect it from damage \fcolorbox{black}{orange}{\rule{0pt}{3pt}\rule{3pt}{0pt}}. In between, wind power generation first increases rapidly with increasing wind speed \fcolorbox{black}{blue}{\rule{0pt}{3pt}\rule{3pt}{0pt}}. Once the maximum wind speed that the wind turbine can convert to power is reached, the power output is usually constant until the wind speed exceeds the cut-out wind speed \fcolorbox{black}{green}{\rule{0pt}{3pt}\rule{3pt}{0pt}}.}
    \label{fig:power-curve}
\end{figure}

As a last step, we relate the insights from the wind speed distributions to multi-decadal wind power forecasts. Wind power generation is often forecasted using hub-height wind speed forecasts with a turbine-specific wind power curve \citep{wang2019approaches} that describes the relationship between wind (speed) and potential wind power generation and is highly non-linear. We apply the Enercon E92/2350 wind power curve,visualized in \Cref{fig:power-curve}. Given the relatively short duration of the main observational datasets we use in this study (around 3-5 years of non-missing data) and our interest in multi-decadal forecasts, we study potential power generation using four $18$ to $34$ year long wind speed datasets.

Additionally, to determine whether the results we find for observational datasets are also applicable to climate model data, we repeat our analysis using data generated by the historical run of the MPI-ESM1-2-LR general circulation model \citep{eyring2016overview} which was the only model with three-hourly data available on the ESGF node \citep{esgf} at the time of research. 
%If $i$ is the $i$'th 10min observation of the original time series we discard all $w_i$ where index $i \mod t \neq 0 $. The $j$'th wind speed value $w_j$ in the shorter simulated instantaneous time-series is then computed as
%\begin{equation}\label{eq:instantaneous}
%    w^t_j = w^{1}_{jt}.
%\end{equation}
%For the data considered this is equivalent to using hourly observations every full hour, three-hourly and six-hourly observations beginning at $0$ am and daily observations every night at $0$ am.  
%apply wind power curve? this could also be an extra section
%aditionally: use one year of data only to show that dataset is large enough (even one year of data is enough so 5 years are more than enough)

\section{Results}
\label{results}
\begin{figure}
    \centering
    \includegraphics[width=0.5\textwidth]{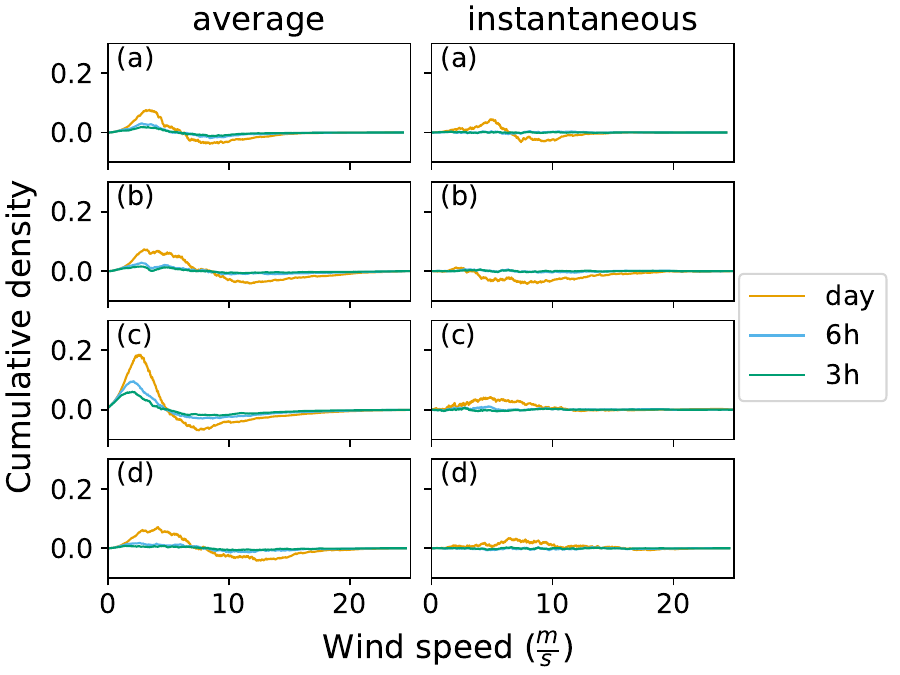}
    \caption{Difference of cumulative densities from the 10min data to the other temporal resolution datasets for average wind speeds (left) and instantaneous wind speeds (right). It can be seen that the averaged wind speeds are visually distinguishable, which is less the case for instantaneous wind speeds, particularly for data with a temporal resolution of six-hourly or higher.}
    \label{fig:cumdens}
\end{figure}

We analyze the distributions of wind speed averages and instantaneous wind speed time series, and find consistent results across all sites investigated: averaging introduces shifts to the wind speed distributions, while three-hourly and six-hourly instantaneous data are usually close to the original.
% variance changes
This is clearly demonstrated in \Cref{fig:cumdens}, with differences between 10min wind speeds and averaged data (left hand column) larger than the differences between 10min wind speeds and instantaneous wind speeds of lower resolution (right hand column). Similar patterns can be observed in the four longer datasets, see \Cref{appendix:cumdens-long}.

\begin{figure}
    \centering
    \includegraphics[width=0.5\textwidth]{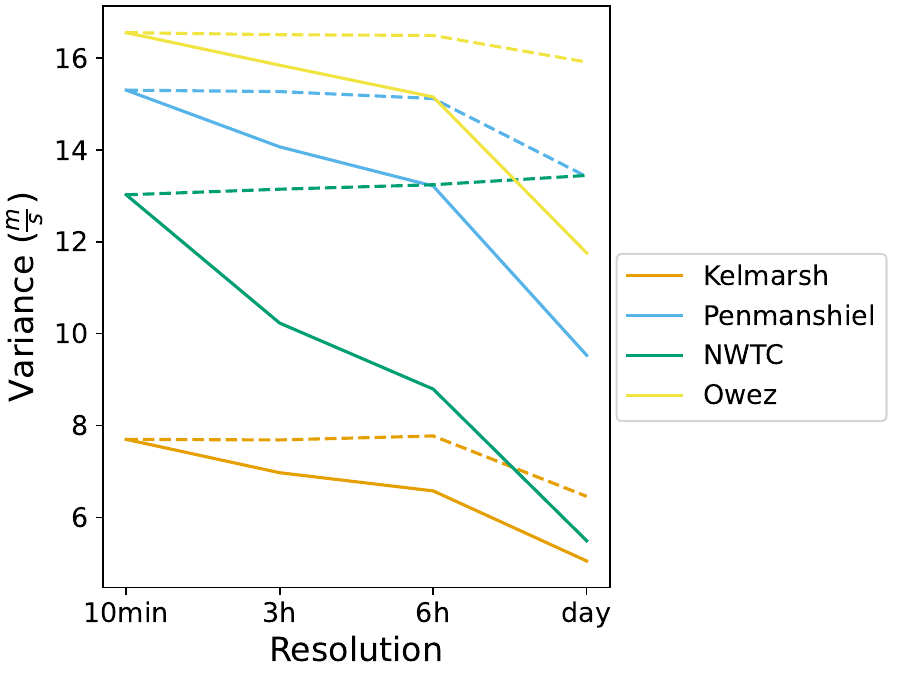}
    \caption{Variances of averaged (solid lines) and instantaneous (dashed lines) values. The variances steadily decrease when wind speeds are averaged and stay close to the 10min variances for the instantaneous three-hourly and six-hourly distributions.}
    \label{fig:variances}
\end{figure}

The impact of averaging can also be seen in the variance of the data: while averaging does not affect the mean it reduces the variance of the averaged time series (\Cref{fig:variances}). In contrast, three-hourly and six-hourly instantaneous values (dashed lines in \Cref{fig:variances}) preserve the variance, with daily instantaneous values preserving substantially more of the variance than daily averages, and at some sites, more than six-hourly averages.

% Compare the cumulative density distributions, statistical significance
\subsection*{Kolmogorov-Smirnov tests}
To quantify whether the difference between the 10min wind speed distribution and the averaged and instantaneous distributions of lower resolution are statistically significant we compute Kolmogorov-Smirnov test statistics of their cumulative density distributions pairwise. The results for the Penmanshiel site are presented in \Cref{tab:KS_penmanshiel_avrg} (for averaged data) and \Cref{tab:KS_penmanshiel_inst} (instantaneous data), where values in \textbf{bold} indicate no statistically significant difference in the distributions, i.e. that using the dataset of lower temporal resolution may retain all information about the wind speed distribution. We observe that all averaged distributions, as well as daily instantaneous distributions, differ significantly from the 10min wind speeds (right-most column), whilst three- and six-hourly instantaneous values are not distinguishable from the 10 min data. The results for all other locations are very similar (see \Cref{appendix:KS_kelmarsh_avrg} to \Cref{appendix:KS_fichtelberg_inst}): in general, averages do not preserve the wind speed distribution, and three- and six-hourly instantaneous values are almost always not statistically distinguishable from the 10 min data.

\begin{table}[H]
    \centering
    \small
    \begin{tabular}{|c|c|c|c|c|c|c|}
    \cline{2-5}
    \multicolumn{1}{c|}{} & daily & six-hourly & three-hourly &  10min \\
    \hline
    daily&$\mathbf{1}$&$1.12\cdot10^{-3}$&$2.94\cdot10^{-5}$& $1.40\cdot10^{7}$ \\ \hline
     six-hourly&&$\mathbf{1}$&$\mathbf{4.41\cdot10^{-1}}$&$5.83\cdot10^{-3}$ \\ \hline
    three-hourly&&&$\mathbf{1}$& $5.83\cdot10^{-3}$ \\ \hline

    \end{tabular}

    \caption{Test statistics of Kolmogorov-Smirnov test for Penmanshiel averages. We reject the hypothesis that the wind speed distributions are equal if $p \leq 5\cdot10^{-2}$. Therefore, only the three-hourly averages are not significantly different from six-hourly averages.}
    \label{tab:KS_penmanshiel_avrg}
\end{table}

\begin{table}[H]
    \centering
    \small`
    \begin{tabular}{|c|c|c|c|c|c|}
    \cline{2-5}
    \multicolumn{1}{c|}{} & daily & six-hourly & three-hourly &  10min \\
    \hline
    daily&$\mathbf{1}$&$3.03\cdot10^{-2}$&$2.34\cdot10^{-2}$& $6.62\cdot10^{-3}$ \\ \hline
    six-hourly&&$\mathbf{1}$&$\mathbf{1.00}$&$\mathbf{9.58\cdot10^{-1}}$ \\ \hline
    three-hourly&&&$\mathbf{1}$& $\mathbf{8.07\cdot10^{-1}}$ \\ \hline

    \end{tabular}
    \caption{Test statistics of Kolmogorov-Smirnov test for Penmanshiel instantaneous data. We reject the hypothesis that the wind speed distributions are equal if $p \leq 5\cdot10^{-2}$ which reveals that only daily instantaneous values are significantly different from all other distributions.}
    \label{tab:KS_penmanshiel_inst}
\end{table}
% and what about subsampling without averaging?
%Additionally, the Kolmogorov-Smirnov test statistics reveal that, in contrast to averaging, many low-resoluted distributions do not differ significantly from their high-resoluted counterpart for temporal resolution at least six-hourly . In all datasets investigated, the distributions using three-hourly data are not significantly different from their underlying 10min distribution. Additionally, six-hourly averages are usually not significantly different from three-hourly or 10min observations. However, the distributions of daily wind speeds mostly differ significantly from 10min and three-hourly wind speeds.

\subsection*{Changes in Weibull parameters}
% But can we qunatify this shift? Model wind speeds as weibull
The cumulative density plots and Kolmogorov-Smirnov tests indicate that wind speed distributions are likely to \textit{change} when wind speeds are averaged and likely to stay similar when measurements are discarded, at least until around six-hourly resolution. The aim of the next steps is to \textit{quantify these changes} by parameterizing the distributions as Weibull distributions.
%and verify that this is a valid parametrization using quantile-quantile plots. It can be observed that the Weibull distribution models most of our quantiles and datasets well. \textcolor{red}{should I be more critical here? Or in the discussion? I talked to a colleague about this, see comment in line 310}
%However, in some cases we observe a slight shift in the right tail of the distribution, i.e. the upper quantiles of the distribution. We therefore compare the log-densities of the Weibull distribution using the fitted parameters to the original data and observe that the shape of the tail distributions is maintained. We therefore consider the three-parameter Weibull distribution a valid fit to our data. 
% So we can quantify this shift! Change of parameters

\begin{figure}
    \centering
    \includegraphics[width=0.5\textwidth]{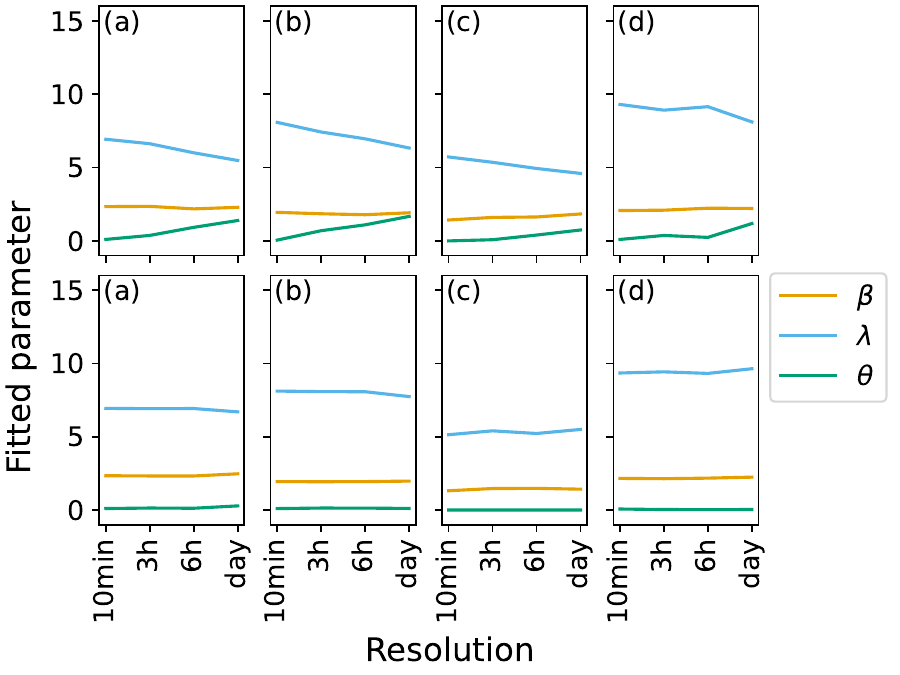}
    \caption{Parameters when data are averaged (top row) or discarded (bottom row) for four different datasets a) Kelmarsh b) Penmanshiel c) NWTC d) Owez. The Weibull parameters change when wind speeds are averaged and stay similar when values are discarded. The corresponding non-parameterized wind speed distributions are visualized in \Cref{appendix:all-kdes} and \Cref{appendix:KDE-instantaneous}.}
    \label{fig:params-weibull}
\end{figure}

\Cref{fig:params-weibull} shows the values of the three Weibull parameters for the different aggregation levels for averaging (top row) and instantaneous (bottom row).
% two-params
%For the two-parameter Weibull distribution the shape parameter $c$ increases in all cases when averaging 10min data to daily values. However, when averaging to sub-daily scales, this parameter increase can not always be observed. In general, there are no large parameter changes below daily aggregation in both parameters. The scale parameter $\lambda$ stays almost constant in all scenarios and is in general much more stable when wind speeds are averaged compared to the shape parameter. 
%three-params
The shape parameter $\beta$ stays approximately constant across all aggregation levels and types, for both averaged and instantaneous data. For averaging, the location parameter $\theta$ increases with higher aggregation levels. This is consistent with the lowest values of the dataset increasing as they are averaged. Conversely, with lower resolution instantaneous data, the lowest values remain similar, leading to similar $\theta$ for all resolutions studied. The scale parameter $\lambda$ decreases with increased averaging length, and remains relatively constant for instantaneous data, consistent with the changes in variance shown in \Cref{fig:variances}. We observe very similar results for $\sim30$ years of data measured at $10m$ height (see \Cref{appendix:weibull-others} and \Cref{appendix:weibull-others-instantaeous}). In addition, two of our observational sites, NWTC and Owez, have wind speed data at multiple heights, ranging from $10m$ to $130m$; the changes in Weibull parameters seen in \Cref{fig:params-weibull} are found at all different heights studied (see \Cref{appendix:obs-heights-rel}).

To test for robustness of our results we also use an MLE to fit a generalized Gamma distribution (see Appendix for details); both Weibull and Gamma distributions are regarded as suitable statistical models for wind speed data \citep[e.g. ][]{mert2015statistical}. We find very similar results (see \Cref{appendix:QQ-plot-Gamma-avrg} to \Cref{appendix:gamma-inst-rel}), with large changes in parameters when averaging data, and relatively small changes for three- and six-hourly instantaneous values. This suggests our conclusions are not sensitive to our choice of parameterization.

\subsection*{Implications for multi-decadal wind power forecasting}
\label{implications}
As introduced in \Cref{introduction}, wind speeds and wind speed variability are subject to interannual variability and climate change. Hence, for multi-decadal wind power forecasts climate models can provide useful information \citep[e.g. ][]{tobin2015assessing}. To understand how our results apply on multi-decadal timescales, we repeat our analysis using data from four sites where multiple decades of $10m$ wind speed observations are available. We also repeat our analysis on climate model output data to determine whether our conclusions are applicable to model data. We extract $10m$ wind speeds from the historical run of the CMIP6 model MPI-ESM1-2-LR for grid points closest to these four longer-term observational sites (locations shown in \Cref{appendix:locations-DWD}). We only use direct output from the model, and thus at daily temporal resolution we do not have instantaneous values, only averages. 

\begin{figure}
    \centering
    \includegraphics[width=0.5\textwidth]{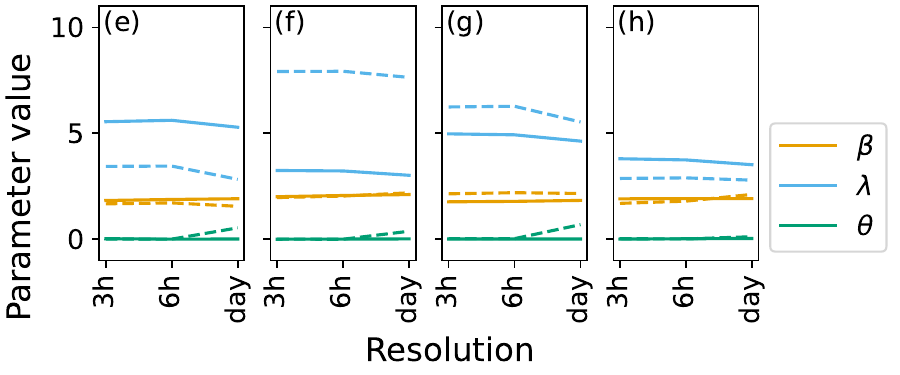}
    \caption{Parameters of Weibull distribution fitted to observational data averages (dashed lines) of the four multiple-decadal sites and to the closest grid points in the CMIP6 MPI-ESM1-2-LR dataset (solid lines). The parameter $\lambda$ decreases in all cases when comparing three-hourly to daily averages, indicating a decrease in variability. The location parameter $\beta$ does not show any consistent trends. $\theta$ increases with daily averages in observational data, but not in the CMIP6 data. }
    \label{fig:cmip6}
\end{figure}

For these multi-decadal datasets the Kolmogorov-Smirnov tests and Weibull parameterization analysis produce results comparable to those for the hub-height observations shown in the previous sections. The parameters of fitted Weibull distributions behave similarly in response to changing temporal resolution, with a decrease of $\lambda$ as temporal resolution decreases indicating a decrease in variability (see \Cref{fig:cmip6}). The climate model data do not show an increase in $\theta$ with decreasing temporal resolution, although for three-hourly and six-hourly averages $\theta$ fitted to the climate model data is very close to the parameters fitted to the observational data.

So far we have only looked at wind speeds and their distributions. However, these are just a proxy for wind power -- our variable of interest -- and wind power depends non-linearly on wind speed. To estimate the power a wind turbine could generate over its lifetime, we apply the Enercon E92/2350 power curve (see \Cref{fig:power-curve}) to the four multi-decadal observational wind speed datasets as well as to the wind speeds from the corresponding closest grid-points in the CMIP6 dataset. We then compare the expected cumulative power generation of the highest available resolution (10min in observations and three-hourly instantaneous values in CMIP6 model) to lower resolutions (three-hourly, six-hourly, and daily averages and instantaneous values). Although the highest available resolution in the CMIP6 model is only three-hourly, our previous analysis shows that these values are closely aligned with 10min data.

% 10min resolution data. Since the highest available resolution output from the MPI-ESM1-2-LR model is three-hourly instantaneous values, and our analysis of the observational data shows that these values are likely closely aligned with 10min averages, we use these three-hourly instantaneous values as the 'truth' for the climate model comparison.
%We compare the cumulative power generation predicted given the observational data of different resolutions to their underlying data of 10min resolution. As three-hourly instantaneous values are the highest resolution at which the CMIP6 date is available and the previous analysesof the observational data suggest that three-hourly instantaneous values are very close to 10min averages, we compare the CMIP6 data aggregations to these. 

\begin{figure}
    \centering
    \includegraphics[width=0.5\textwidth]{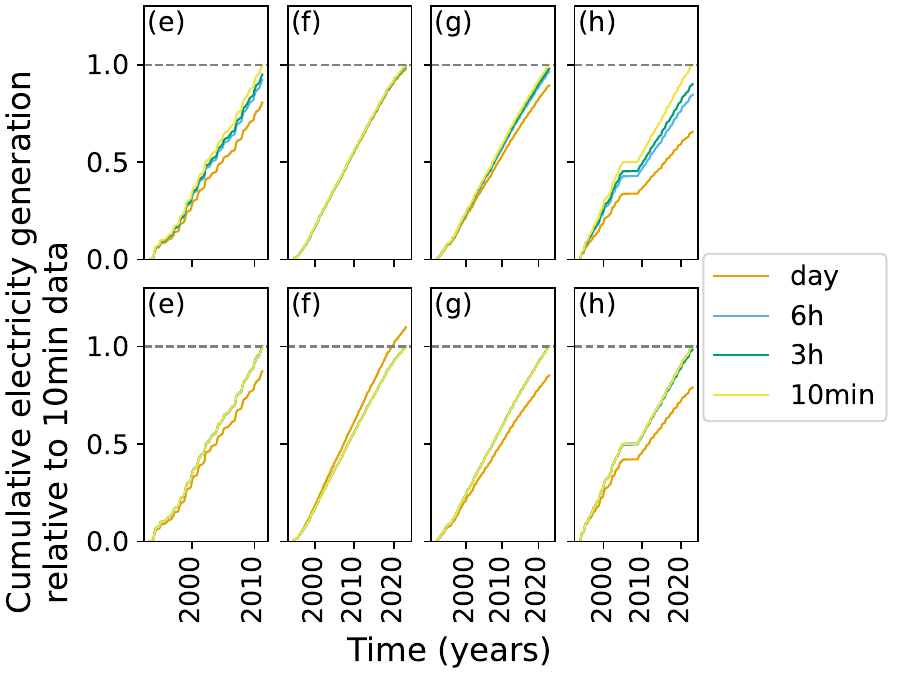}
    \caption{Cumulative power generation of daily, six-hourly and three-hourly values relative to the total 10min wind power generation computed by feeding the wind speeds into a power curve. Top row: Average values underestimate wind power generation -- the longer the forecast the higher the error. Bottom row: Instantaneous values every three or six hours only introduce minor errors. Daily data is in no case a good proxy. The \textit{power generation gap} at the Fichtelberg site between $\sim 2005$ and $2010$ stems from various missing observations. Absolute errors are reported in \Cref{appendix:errors}.}
    \label{fig:power-real}
\end{figure}

 \Cref{fig:power-real} shows the expected cumulative power generation using wind speed observations of different resolutions. We show this as a fraction of the total power generation achieved when applying the wind power curve to the 10min observational wind speed data and integrating over the whole time period. The dotted gray line shows 100\% and thus if the expected cumulative power generation reaches this threshold without overshooting we consider the change introduced by the temporal resolution to be small.
% relative to the total power generated over the time period using the 10min data.  This total power generation is shown by the dotted gray line and will be reached by the expected cumulative power geenration  reaching a value of 1, shown by the dotted gray line, by the end of the time series). We consider the change in temporal resolution to have minimal impact if the line reaches that threshold of 1.0 without overshooting. 
The top row of \Cref{fig:power-real} shows that averaging values, particularly to daily, but even to three- or six-hourly at some locations, can lead to relatively large errors in estimated power generation, with errors of up to -34.48\% (daily average), -15.45\% (six-hourly average), and -10.06\% (three-hourly average). 
Conversely, three-hourly and six-hourly instantaneous values, shown in the bottom row, reveal very similar results to the 10min data, with errors less than 2\%. %, with the largest differences -21.03\% (daily instantaneous), -1.57\% (six-hourly instantaneous), and +0.29\% (three-hourly instantaneous). 

\begin{figure}
    \centering
\includegraphics[width=0.5\textwidth]{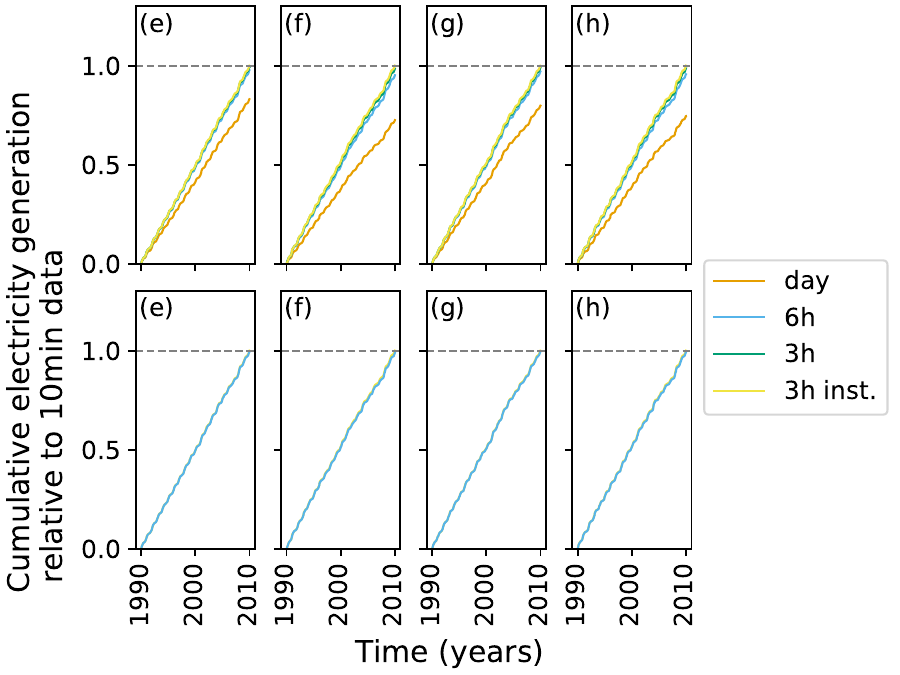}
    \caption{Cumulative power generation of daily, six-hourly and three-hourly values relative to three-hourly instantaneous (3h inst.) wind power generation at the four CMIP6 locations closest to e) Aachen, f) Zugspitze, g) Boltenhagen, h) Fichtelberg computed by feeding the wind speeds into a power curve. Top row: Average values underestimate wind power generation. Bottom row: Instantaneous values (note that daily instantaneous values were not available as direct output, and so are not included). Using six-hourly instantaneous values introduces only minor errors relative to three-hourly instantaneous values.}
    \label{fig:power-cmip}
\end{figure}

\Cref{fig:power-cmip} demonstrates that very similar results are found using the output from the MPI-ESM1-2-LR global climate model, with results given relative to the total amount of power generated using three-hourly instantaneous values. %The largest differences to three-hourly instantaneous data are -27.44\% (daily average), -4.79\% (six-hourly average), -1.51\% (three-hourly average) and -1.1\% (six-hourly instantaneous). 
In all cases six-hourly instantaneous values are closest to three-hourly instantaneous values, followed by three-hourly averages and six-hourly averages; daily averages differ substantially.

Reducing temporal resolution leads to an underestimation of expected power generation at all sites studied except for daily instantaneous values at Zugspitze (\Cref{fig:power-real} bottom row, site (f)), a site with relatively high mean wind speeds and high wind speed variance, situated at a high elevation above sea level ($2956m$). This is very likely a function of the particular wind turbine power curve we have chosen. It is important to note that in all cases, both with observational data and climate projections, the difference in power generation compared to higher-resolution data increases with an increasing forecast horizon and does not average out - the shift in the wind distribution leads to a systematic error in wind power estimation. In general, averaging tends to result in an underestimation of expected power generation, while discarding values has only minor impacts.
 
%However, in all cases -- real data and climate projections -- the difference in power generation compared to higher resoluted data increases with increasing forecast horizon and does not average out. 
% cmip6 here 
% show one example of different power generation using different resolutions

% in numbers: from 5 to 8
% avrg
% daily: 80.56%, 97,98%, 89.35%, 65.52%
% 6h: 94.90%, 99.50%, 96.33%, 84.55%
% 3h: 92.12%, 99,11%, 97.88%, 89.94%
% inst 
% daily: 87,25%, 109.90%, 85.11%, 78,97%
% 6h: 99.56%, 100.06%, 100.19%, 98.43%
% 3h: 100.25%, 100.07%, 100.07%, 100.29%

% in numbers from 1 to 8 
% day 80.84%, 80.97%, 60.18%, 91.72%, 83.14%, 72.56%, 79.85%, 74.59%
% 6hr avrg 97.51%, 97.62%, 93.22%, 99.13%, 97.75%, 95.21%, 97.06%, 95.70%
% 3hr avrg 99.29%, 98.68%, 97.83%, 99.76%, 99.35%, 98.49%, 99.11%, 98.68%
% 6hr inst 99.66%, 99.83%, 100.03%, 99.96%, 99.53%, 98.90%, 99.44%, 99.21%

\section{Discussion}
\label{discussion}

% bring the climate context: We'll not get 10min resolution but it's also not necessary
Global climate models simulate climate dynamics physically using partial differential equations. Their long forecast horizons make them computationally expensive with large storage space needs and their output is usually provided with a temporal resolution ranging from three hours to one month, either as instantaneous or averaged values. Using hub-height wind speed observations, this study investigates how the temporal resolution of wind speed data affects the wind speed distributions. Using multi-decadal observational datasets at 10$m$ height we study the corresponding potential power generation and which time resolution is actually necessary. 

% Instantaneous wind speeds are closer to the 10min distribution. Summarize overall results.
Using hub-height data we find that in all cases investigated, three-hourly and six-hourly instantaneous observations well-preserve the underlying 10min wind speed distribution, whilst three- and six-hourly averaging leads to distributional shifts. 
%We describe these shifts in terms of parameter changes of fitted Weibull distributions (compare \Cref{fig:params-weibull}), finding a decrease in variability when wind speeds are averaged. 
However, whether a significant wind speed distribution shift results in a significant change in wind power generation depends on the turbine and its corresponding power curve. Our results, using an example turbine power curve, indicate that the differences in wind distribution highlighted in this study can lead to accumulating errors when power generation is forecasted for years to decades (compare \Cref{fig:power-real} and \Cref{fig:power-cmip}). 
% representativesness, 3 years vs 30 years vs 1 year

% The parameter shifts that are persistant when averaging are an indicator when wind power generation is over-\underestimated: Power curve is needed for this distinction.
%Whether a significant wind speed distribution shift results in a significant change in wind power generation depends on the turbine and its corresponding power curve. Our results indicate that the differences found in this study can lead to accumulating errors when power generation is forecasted for years to decades (compare \Cref{fig:power-real} and \Cref{fig:power-cmip}). While the Enercon power curve we use in this study has the characteristic shape of any modern horizontal wind turbine power curve, it is not necessarily representative. Furthermore, for site assessment, wind speeds are usually transformed to hub-height which is non-trivial \citep[e.g. ][]{banuelos2010analysis}. Based on our results, we argue that modelling wind speed \textit{distributions} correctly is what we should aim for in multi-decadal wind power forecasting, rather than average wind speed.

% climate projections
Our results are consistent across different observational sites and a GCM. We can therefore give two suggestions when working with wind speed projections for wind power modelling. First, instantaneous wind speed projections should be preferred over wind speed averages, as sub-sampling wind speed data introduces relatively minor errors in contrast to averaging wind speeds, where we observe a characteristic distributional shift. This shift associated with averaging data indicates that we might consistently over- or underestimate wind power generation. Second, instantaneous wind speed projections of six or three hours suffice, whilst daily data may be too low resolution, even with instantaneous values. For our sites temporal downscaling or either three- or six- hourly data is unlikely to provide substantial improvement in accuracy. For example, instantaneous observational wind speeds with a three-hourly temporal resolution lead to errors of less than $0.29\%$, with six-hourly data leading to errors of less than $1.57\%$. The experiments conducted using climate model data support these results. This knowledge can be useful to reduce storage space almost loss-free and decrease computational complexity in further applications using the data.

The primary shortcomings of our investigations include a potential lack of generalizability across turbines, sensitivity to our underlying highest temporal data resolution, and uncertain transferability to other climate models. More specifically, as wind speed data measured at hub-height is often confidential, we are limited to relatively few datasets. Furthermore, for site assessment, wind speeds are usually transformed to hub-height which is non-trivial \citep[e.g. ][]{banuelos2010analysis}. However, the close agreement of results from stations across a range of different locations, including different local geography (off-shore, on-shore, different altitudes and local topography), suggests that our results likely hold for the majority of locations. For 7 out of the 8 locations studied, using daily instantaneous or average values underestimates the power generation (see \Cref{fig:power-real}), while six-hourly values are good approximations. For one station, however, the daily instantaneous data is an overestimate; understanding the conditions under which daily data over- or under-estimates the power generation would be useful for studies in which only daily data is available. In addition, while the Enercon power curve we use in this study has the characteristic shape of any modern horizontal wind turbine power curve, it is not necessarily representative of the turbines at a particular site.

%10min observations 
Regarding the sensitivity to the underlying data resolution, we use data in 10min resolution, hence variability on shorter time scales -- often associated with turbulence \citep[e.g. ][]{stull1988introduction} -- is not preserved. However, higher temporal resolution data is rarely available or used in wind power forecasting \citep{tawn2022review}, \citep{effenberger2022collection}, and thus we assume that this is a minor issue. This claim is also supported by \cite{lopez2021effects} who investigate wind power spectra \citep{van1957power} and find only small differences between power production given wind speeds of different resolution between 1min and 6h. 

Lastly, climate projections are characterized by different pathways that describe anthropogenic climate change \citep{eyring2016overview}. This makes handling the data cautiously even more important, especially in wind power forecasting, where non-linearly dependent wind speed projections are often used as a proxies for power generation. Our results using one historical CMIP6 run indicate that changes in observational wind speed distributions for different temporal resolution data can be seen in data from climate models as well. However, future research has to investigate the sensitivity of these results to different climate models. %the spatial resolution of climate models, as well as model differences in underlying dynamics and parameterizations of boundary layer turbulence.

\section{Conclusion}
\label{conclusion}
Wind power generation depends non-linearly on wind speeds. For multi-decadal wind power forecasts in the order of years to decades, small changes in modelling the wind speed distributions can result in large systematic errors in power estimation, with absolute errors increasing with longer forecast horizon. Using hub-height wind speed observations and climate model output data, this study investigates how the temporal resolution of wind speed data affects wind speed distributions and corresponding potential power generation. 

We show that instantaneous wind speeds of lower resolution more accurately represent the underlying distribution of higher resolution data when compared to averaged wind speeds. Three- and six-hourly instantaneous values preserve the wind speed distribution of 10min wind speed averages well. Small changes in the wind speed distribution, through averaging or using daily data, has significant impacts on the estimated wind power generation of a turbine over its lifetime. These results hold true across several observational sites and a global climate model. Based on our results, we argue that modelling wind speed \textit{distributions} correctly is what we should aim for in multi-decadal wind power forecasting.

\input{acknowledgements}

\bibliographystyle{plainnat}
\bibliography{bib}
\newpage
\begin{appendices}

\counterwithin{figure}{section}
\counterwithin{table}{section}
\input{appendix}

\end{appendices}
\end{document}

%% file: acknowledgements.tex
\section*{Acknowledgements}
This study was funded by the Deutsche Forschungsgemeinschaft (DFG, German Research Foundation) under Germany's Excellence Strategy – EXC number 2064/1 – Project number 390727645 and the Athene Grant of the University of Tübingen. The authors thank the International Max Planck Research School for Intelligent Systems (IMPRS-IS) for supporting Nina Effenberger, and the Natural Sciences and Engineering Research Council of Canada (NSERC) [RGPIN-2020-05783] for supporting Rachel H. White. We acknowledge support by Open Access Publishing Fund of University of Tübingen. Open Access funding enabled and organized by Projekt DEAL. WOA Institution: Eberhard Karls Universitat Tubingen Consortia Name : Projekt DEAL. Nina wants to thank Roland Stull and the Weather Forecast Research Team at UBC for (intellectual) resources!

%% file: appendix.tex
\section{Supplementary Material}
\label{appendix}
        
In the following we present results from additional datasets. Unless specified otherwise, we assign the following abbreviations to the different datasets: a) Kelmarsh, b) Penmanshiel, c) NWTC, d) Owez,e) Aachen, f) Zugspitze, g) Boltenhagen, h) Fichtelberg. They can be found in the top-left corner of all plots. 

    \begin{table}[H]
\small
    \centering
    \begin{tabular}{
    |p{0.09\textwidth}|
    >{\centering\arraybackslash}p{0.11\textwidth}|>{\centering\arraybackslash}p{0.08\textwidth}|>{\centering\arraybackslash}p{0.08\textwidth}|>
    {\centering\arraybackslash}p{0.08\textwidth}|>
    {\centering\arraybackslash}p{0.09\textwidth}|>{\centering\arraybackslash}p{0.09\textwidth}|}
    \hline
         Name & Lat, Lon & Temporal resolution in $min$& Observation period& Elevation in $m$ & Mean wind speed (variance) in $\frac{m}{s}$ & Data source\\ \hline
         Aachen& 50.78, 6.09 & 10 & 29.04.1993-31.03.2011 & 202.00 & 3.06 (4.19) &\cite{DWD}\\ \hline
         Boltenhagen &54.00, 11.19& 10&11.07.1991-31.12.2022 & 15.00 & 5.54 (8.09)& \cite{DWD} \\ \hline
         Fichtelberg &49.98, 11.84& 10 &16.12.1993-31.12.2022& 654.50 & 2.58 (2.91)&\cite{DWD} \\ \hline
         Zugspitze & 47.42, 10.98 & 10 &31.07.1994-31.12.2022& 2956.00 & 7.00 (15.86)&\cite{DWD} \\ \hline
    \end{tabular}
    \caption{Static data of the four longer-term 10m meteorological measurement sites in Germany. None of the stations was moved during the time period evaluated. }
    \label{appendix:hub-heights-appendix}
\end{table}
\begin{figure}[H]
    \centering    
    \includegraphics[width=0.5\textwidth]{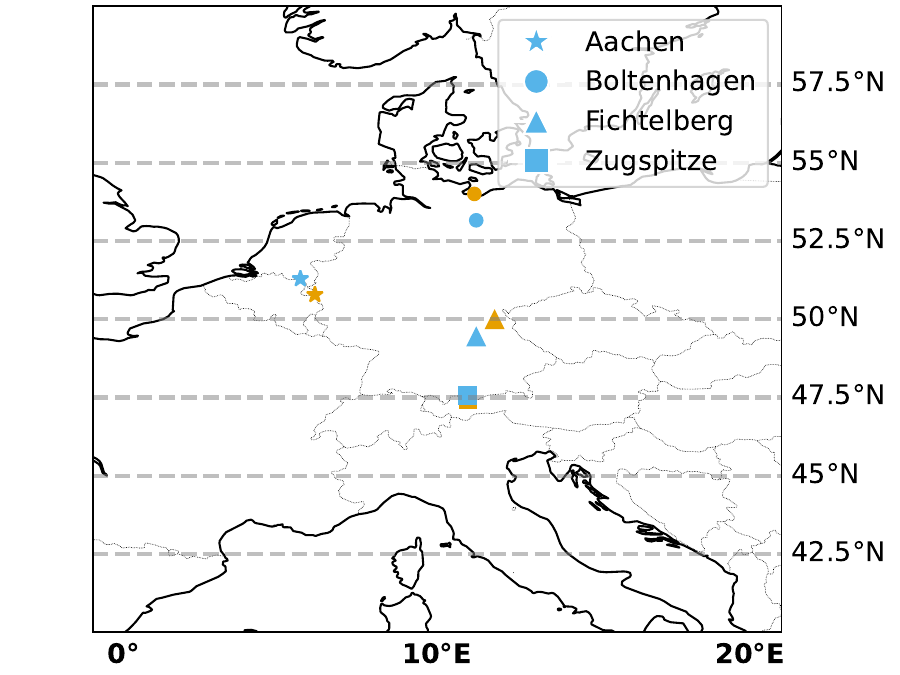}
    \caption{Locations of German weather stations that serve as longer datasets in blue \fcolorbox{black}{bluefull}{\rule{0pt}{3pt}\rule{3pt}{0pt}}. The corresponding CMIP6 locations we investigate for \Cref{fig:cmip6} and \Cref{fig:power-cmip} are marked in orange \fcolorbox{black}{orangefull}{\rule{0pt}{3pt}\rule{3pt}{0pt}}.}
    \label{appendix:locations-DWD}
\end{figure}

\begin{figure}[H]
    \centering
\includegraphics[width=\textwidth]{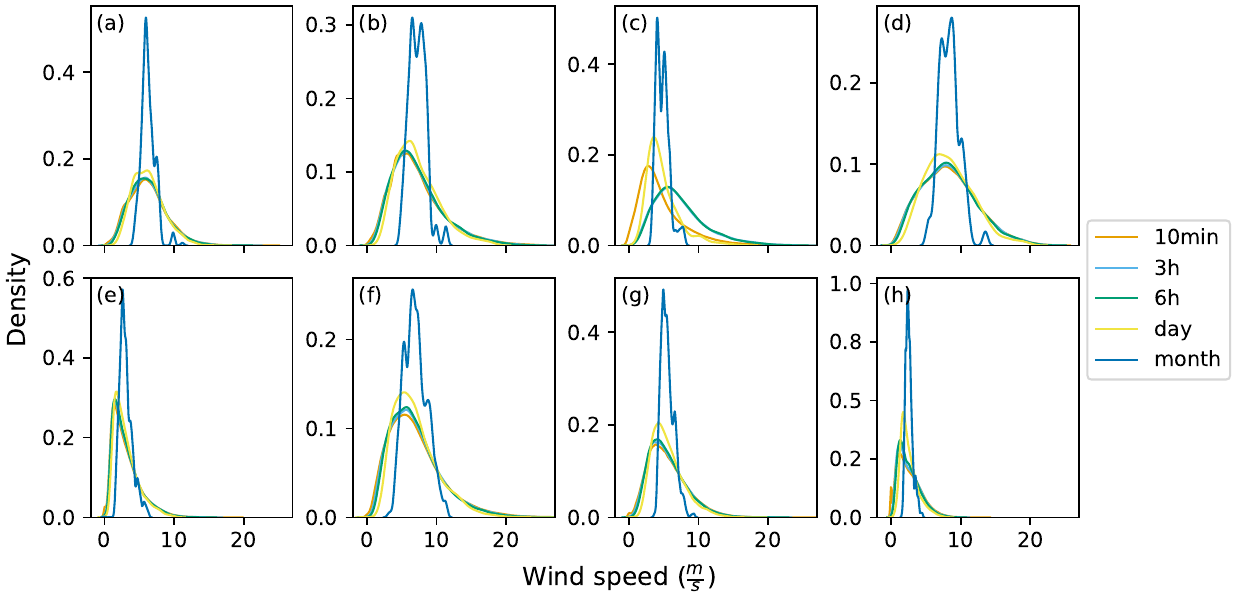}
    \caption{Kernel density estimations of averaged wind speeds. The distributions of 10min, three-hourly, six-hourly and daily data look similar to a Weibull distribution; monthly data are not Weibull distributed.}
    \label{appendix:all-kdes}
\end{figure}

\begin{figure}[H]
    \centering
\includegraphics[width=\textwidth]{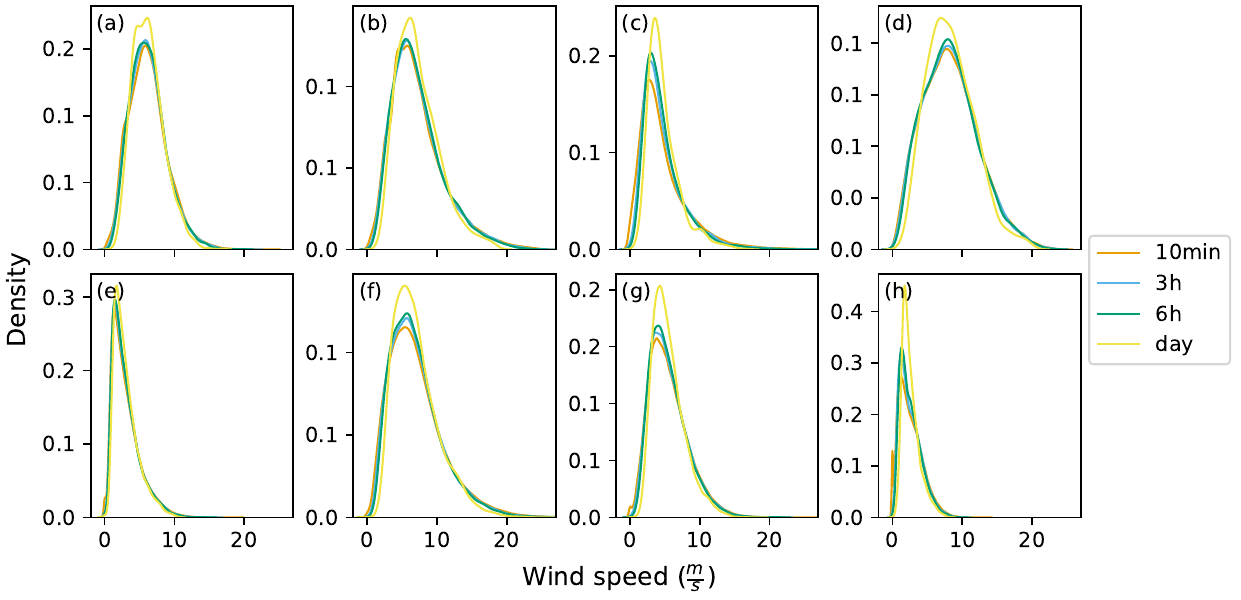}
    \caption{Kernel density estimations of averaged wind speeds of different resolutions excluding monthly data.}
    \label{appendix:KDE-average}
\end{figure}

\begin{figure}[H]
    \centering
    \includegraphics[width=\textwidth]{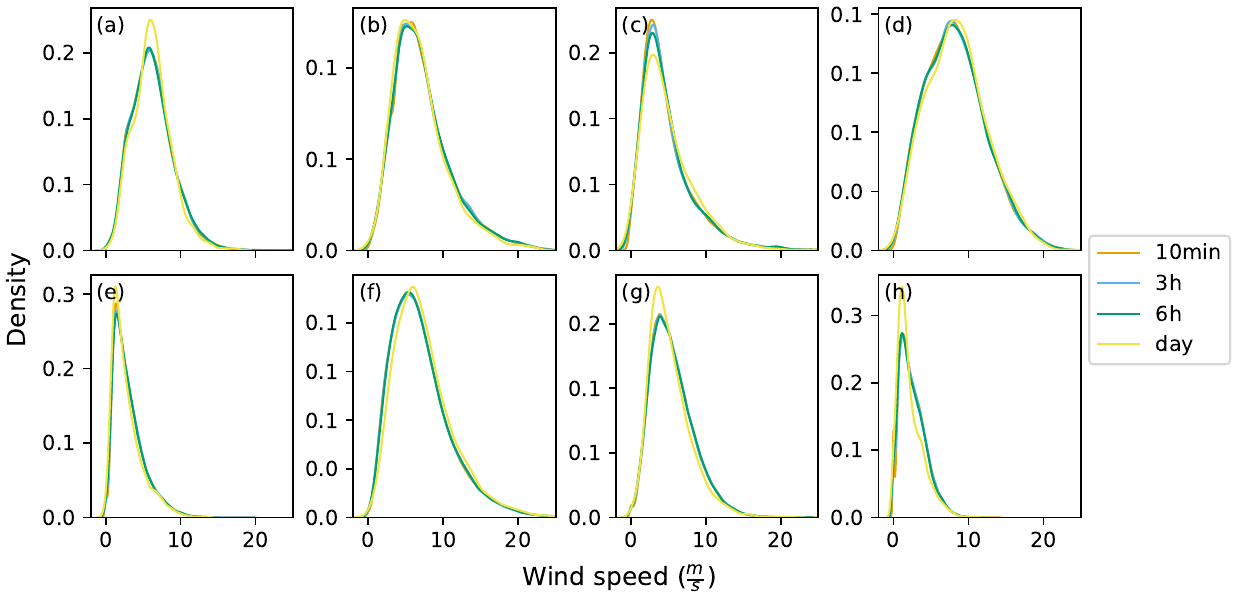}
    \caption{Kernel density estimations of instantaneous wind speeds of different resolutions. Data at most sites look approximately Weibull distributed.}
    \label{appendix:KDE-instantaneous}
\end{figure}

\begin{figure}
    \centering
\includegraphics[width=0.9\textwidth]{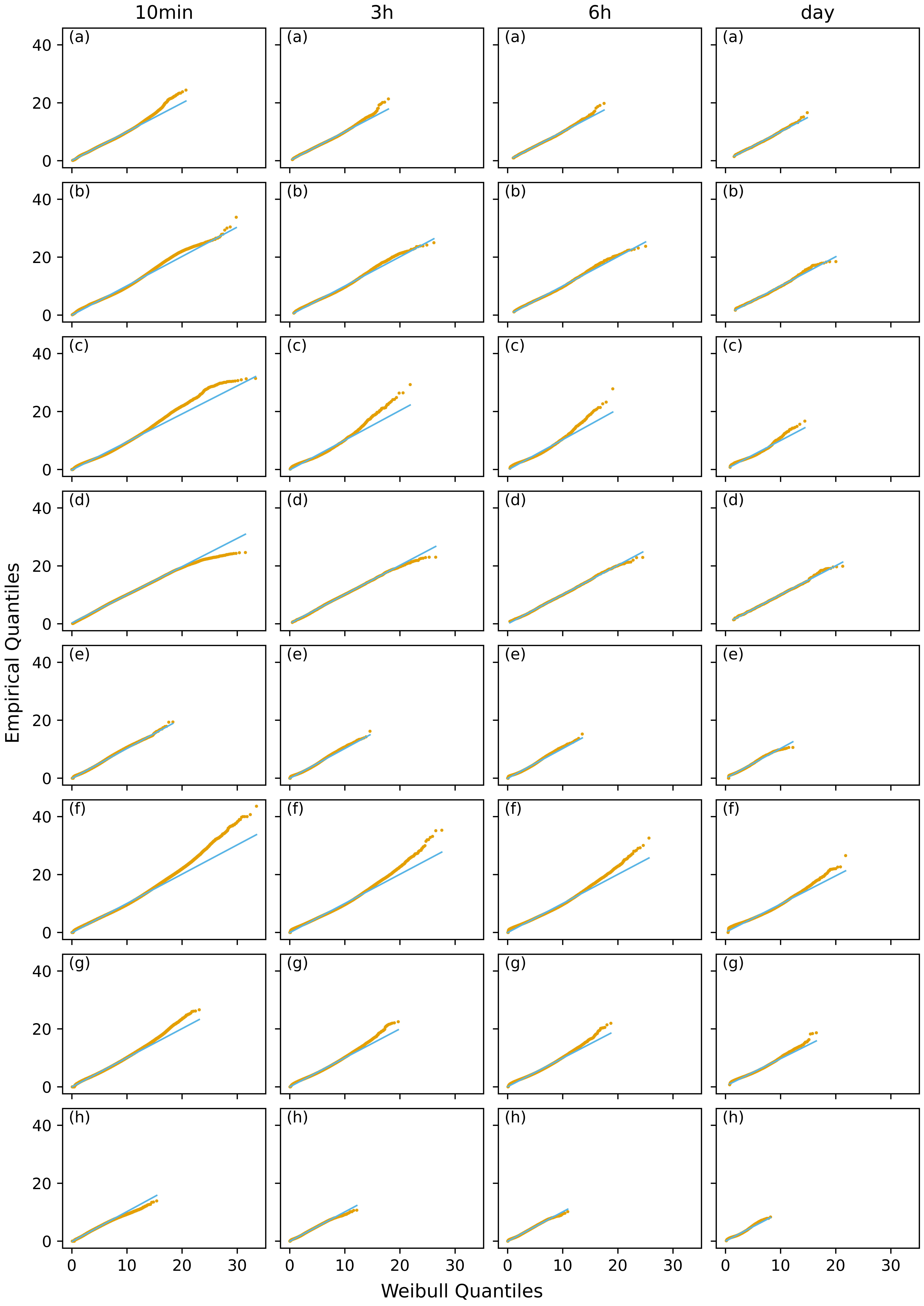}
    \caption{QQ-plots of averaged wind speeds and fitted Weibull distributions. From left to right the data resolution is 10min, three-hourly averages, six-hourly averages and daily averages.}
    \label{appendix:QQ-plot-Weibull-avrg}
\end{figure}

\begin{figure}
    \centering
\includegraphics[width=0.9\textwidth]{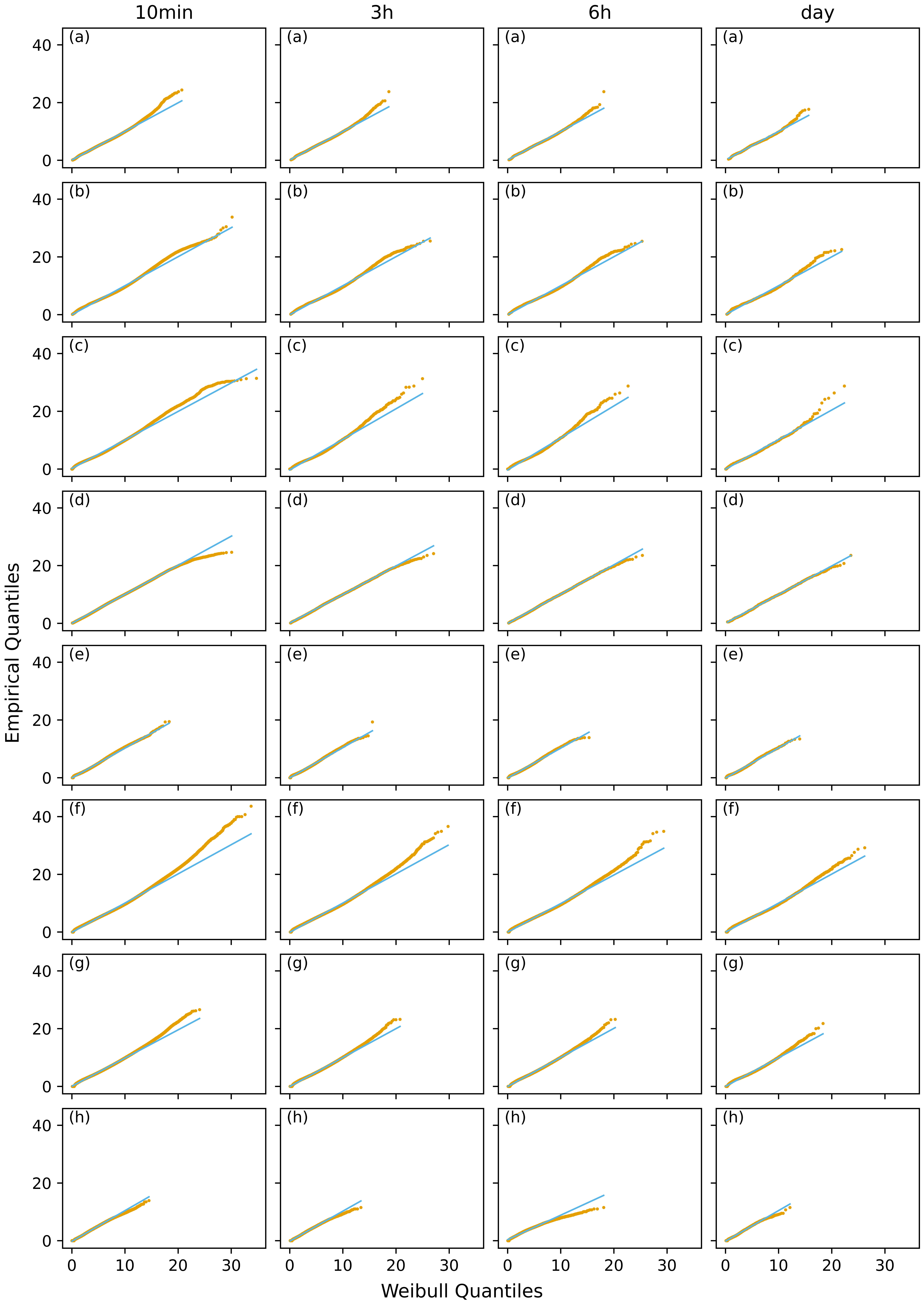}
    \caption{QQ-plots of instantaneous wind speeds and fitted Weibull distributions. From left to right the data resolution is 10min, three-hourly averages, six-hourly averages and daily averages.}
    \label{appendix:QQ-plot-Weibull-inst}
\end{figure}

\begin{figure}[H]
    \centering
    \includegraphics[width=0.5\textwidth]{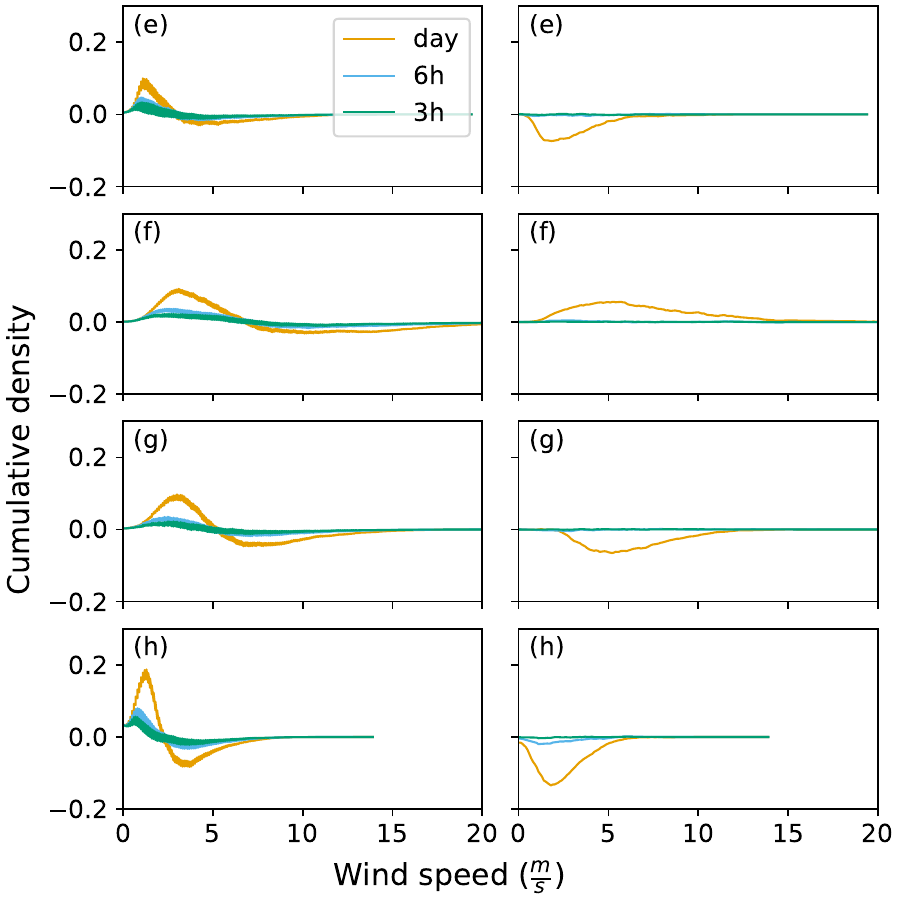}
    \caption{\Cref{fig:cumdens} for longer datasets. Difference of cumulative densities from the 10min data to the other temporal resolution datasets for average wind speeds (left) and instantaneous wind speeds (right). It can be seen that the averaged wind speeds are visually distinguishable, which is less the case for instantaneous wind speeds, particularly for three- or six-hourly data.}
    \label{appendix:cumdens-long}
\end{figure}

\begin{figure}[H]
\begin{subfigure}{.5\textwidth}
    \centering
    \includegraphics[width=\textwidth]{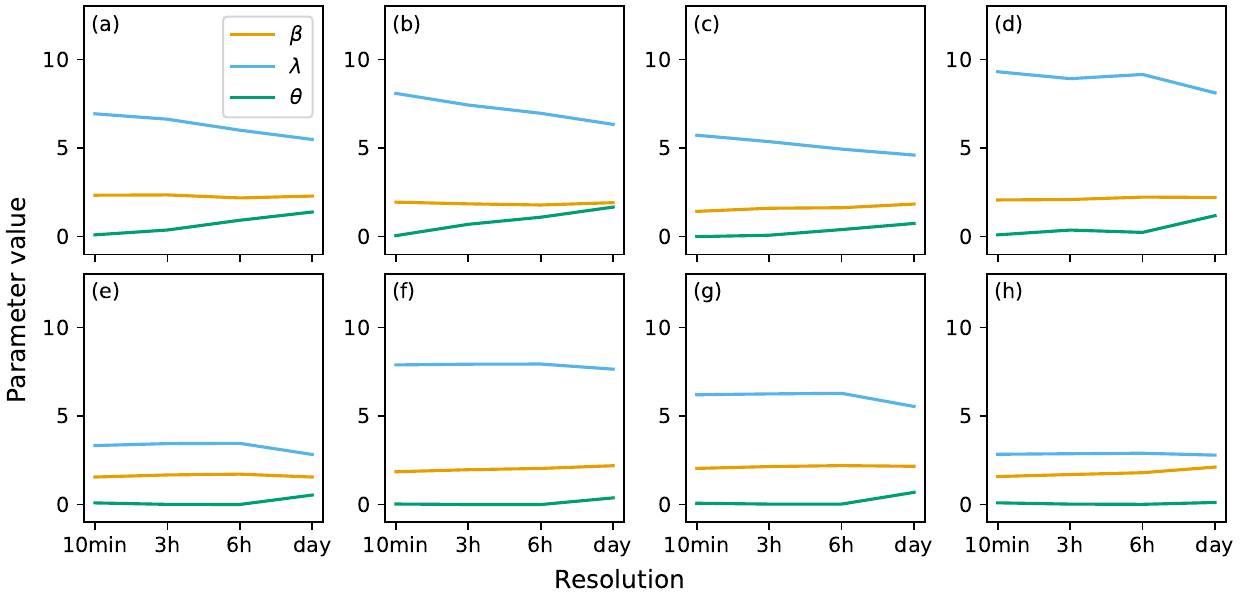}
    \caption{Absolute trends.}
\end{subfigure}
\begin{subfigure}{.5\textwidth}
    \centering
    \includegraphics[width=\textwidth]{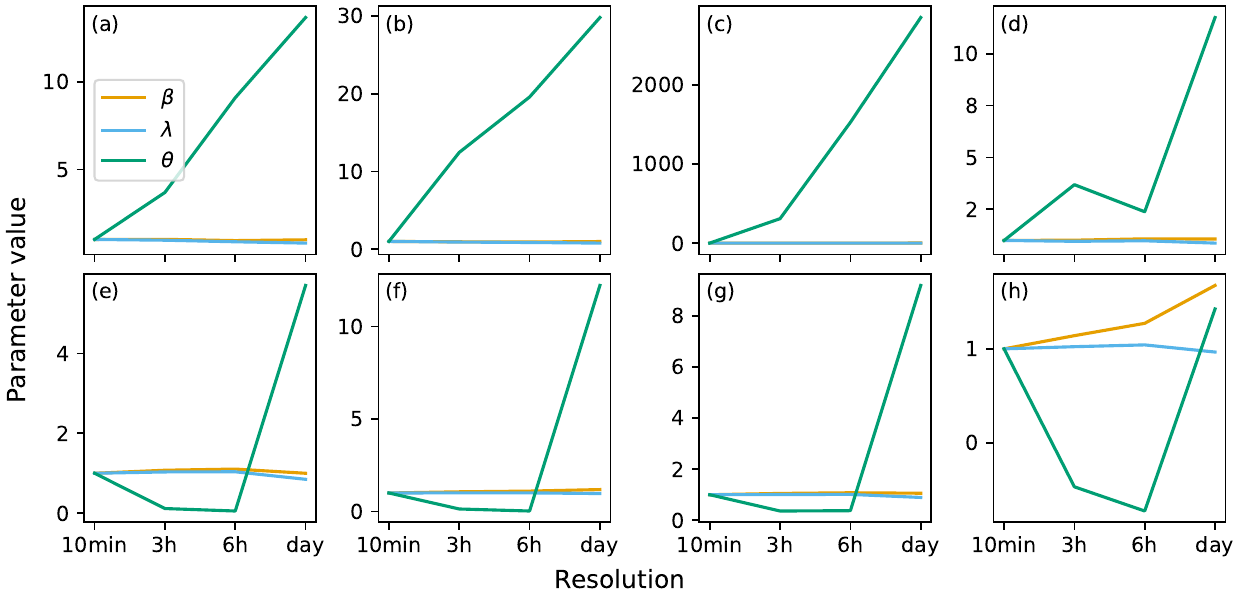}
    \caption{Relative trends.}
\end{subfigure}
\caption{Weibull parameter trends when fitted to averaged wind speed distributions.}
    \label{appendix:weibull-others}
\end{figure}

\begin{figure}[H]
\begin{subfigure}{.5\textwidth}
    \centering
    \includegraphics[width=\textwidth]{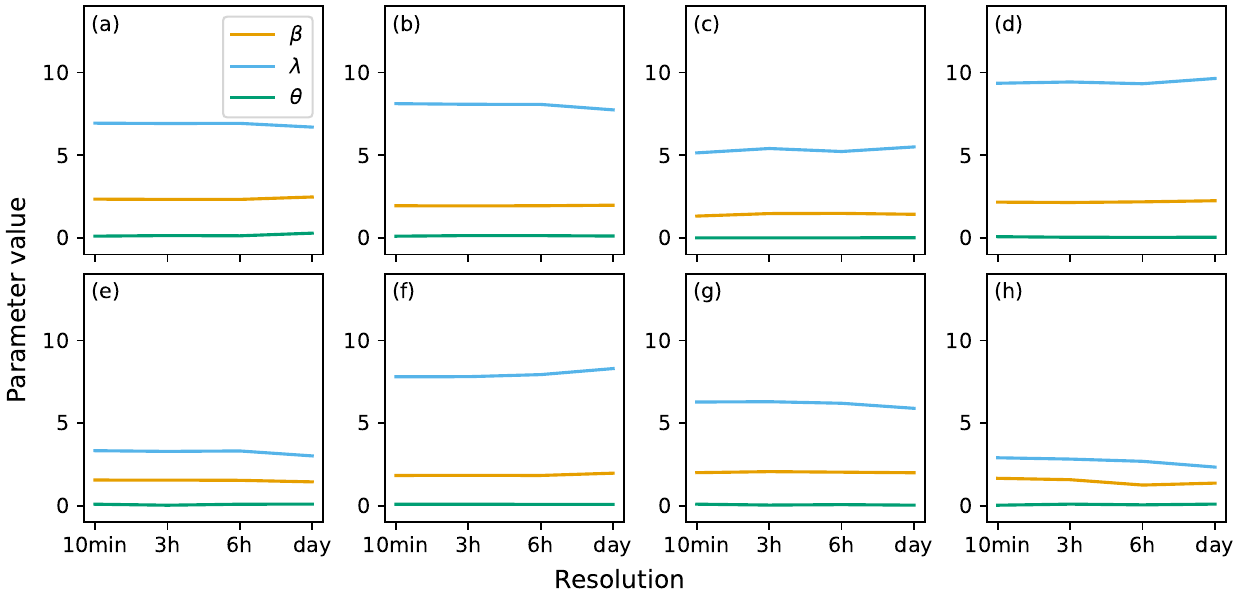}
    \caption{Absolute trends.}
\end{subfigure}
\begin{subfigure}{.5\textwidth}
    \centering
    \includegraphics[width=\textwidth]{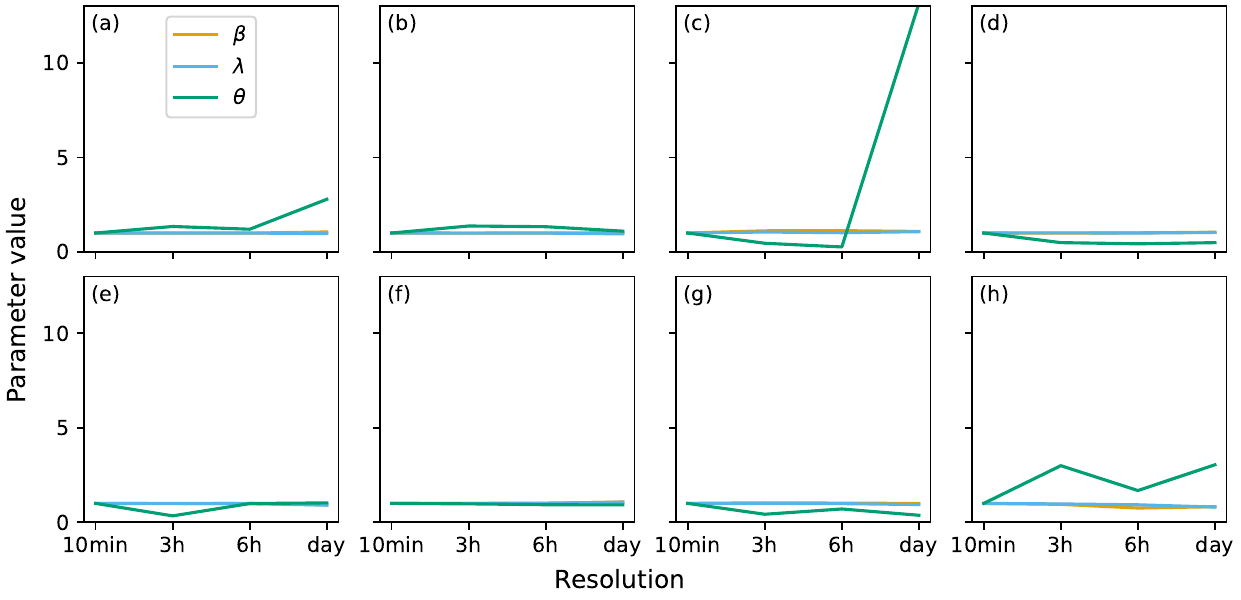}
    \caption{Relative trends.}
\end{subfigure}
\caption{Weibull parameter trends when fitted to instantaneous wind speed distributions.}
\label{appendix:weibull-others-instantaeous}
\end{figure}

\begin{figure}[H]
\begin{subfigure}{.5\textwidth}
    \centering
    \includegraphics[width=\textwidth]{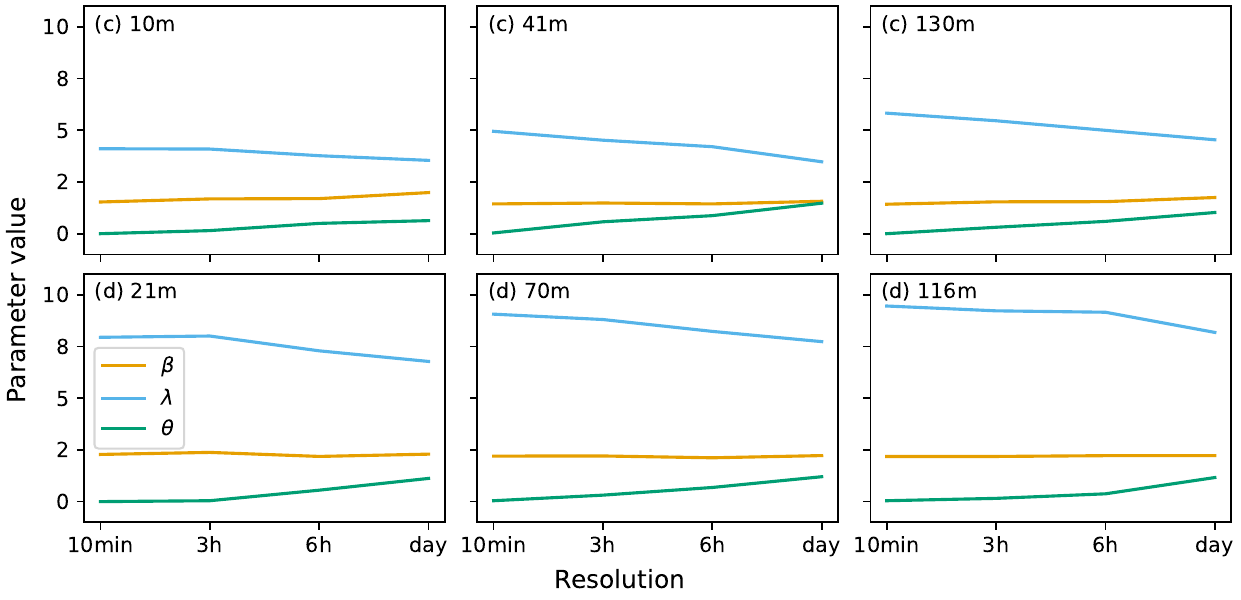}
    \caption{Parameter trends fitted to averaged wind speed distributions.}
\end{subfigure}
\begin{subfigure}{.5\textwidth}
    \centering
    \includegraphics[width=\textwidth]{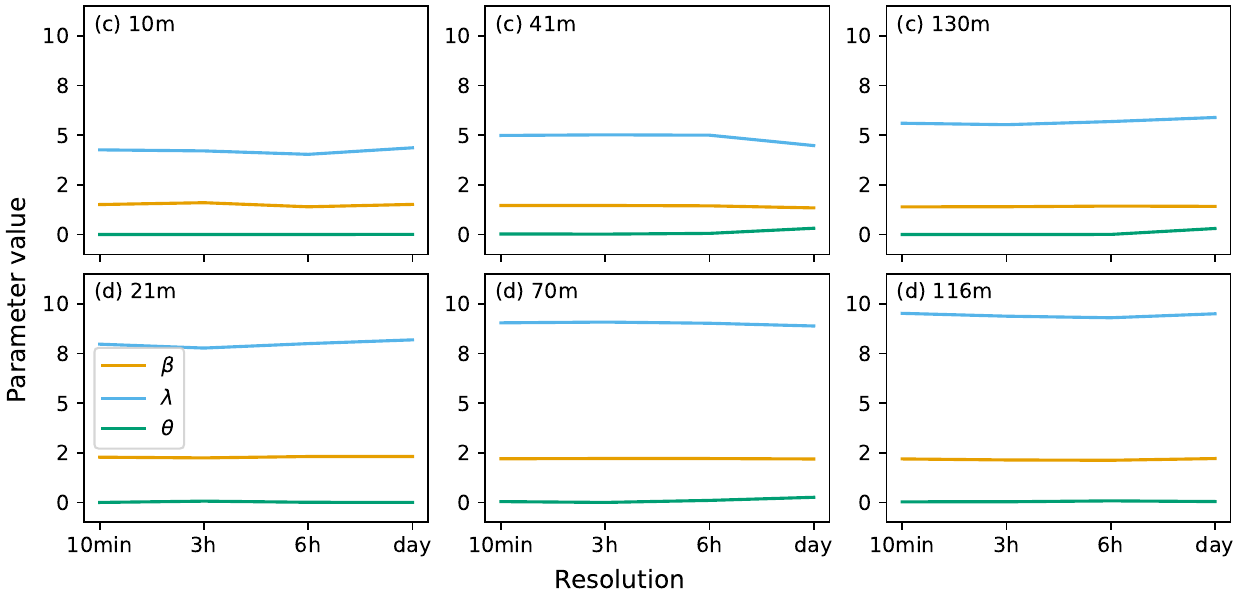}
    \caption{Parameter trends fitted to instantaneous wind speed distributions.}
\end{subfigure}
\caption{Weibull parameter trends of different observations heights for the tall towers NWTC and Owez. }
\label{appendix:obs-heights-rel}
\end{figure}

To test for robustness of our results we also use MLE to fit a generalized Gamma distribution 
\begin{equation}
    f(w; a,d,p)= \frac{\frac{p}{a^d}x^{d-1}\exp(-(\frac{x}{a})^p)}{\Gamma(\frac{d}{p})}
\end{equation}
with 
\begin{equation}
    \Gamma(z)= \int_0^\infty t^{z-1} \exp(-t)dt
\end{equation}
instead of fitting a Weibull distribution. Both distributions were regarded to be suitable statistical models in earlier studies \citep[e.g. ][]{mert2015statistical}. 

\begin{figure}[H]
    \centering
\includegraphics[width=0.9\textwidth]{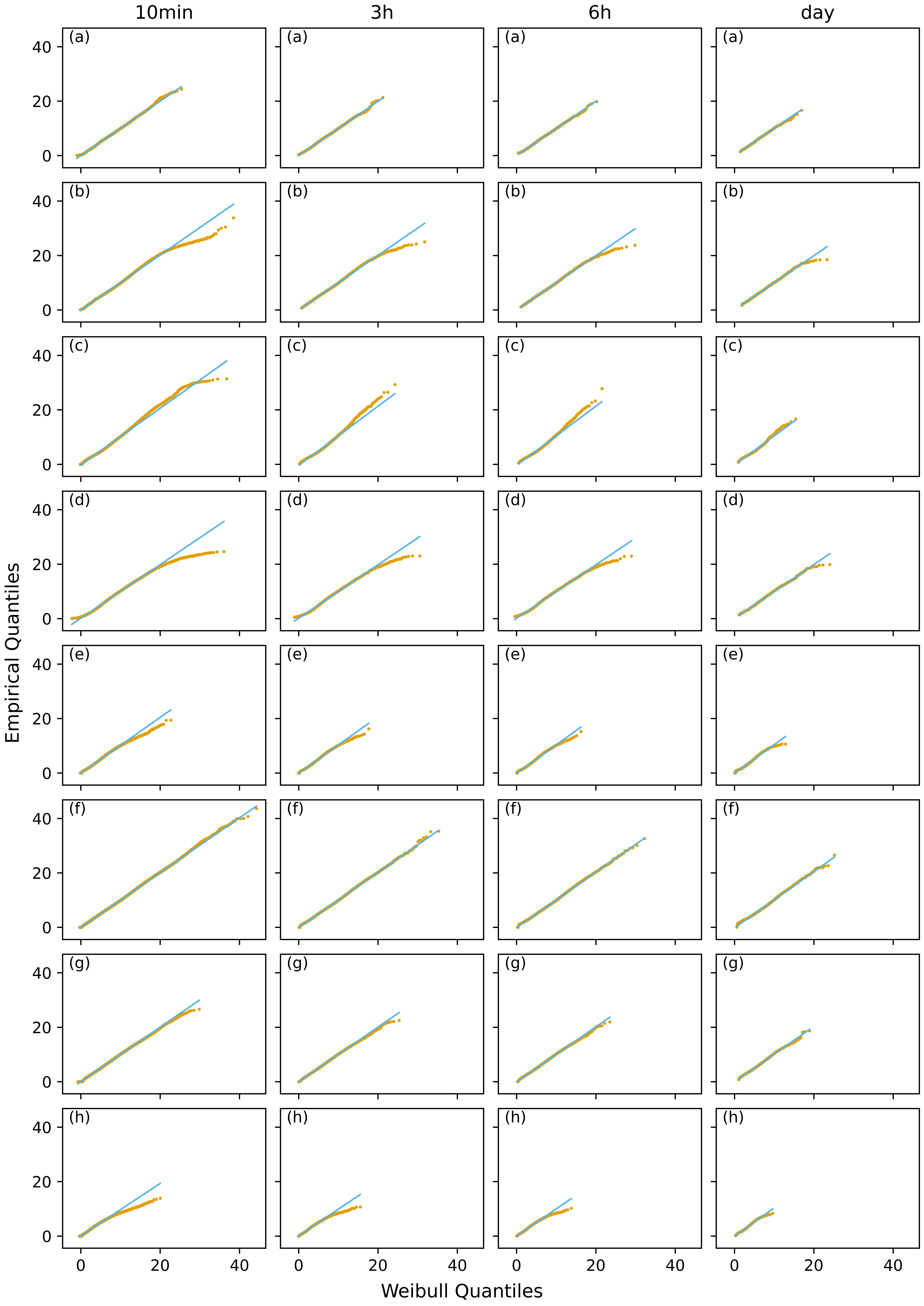}
    \caption{QQ-plots of averaged wind speeds and fitted Gamma distributions. From left to right the data resolution is 10min, three-hourly averages, six-hourly averages and daily averages.}
    \label{appendix:QQ-plot-Gamma-avrg}
\end{figure}

\begin{figure}[H]
    \centering
\includegraphics[width=0.9\textwidth]{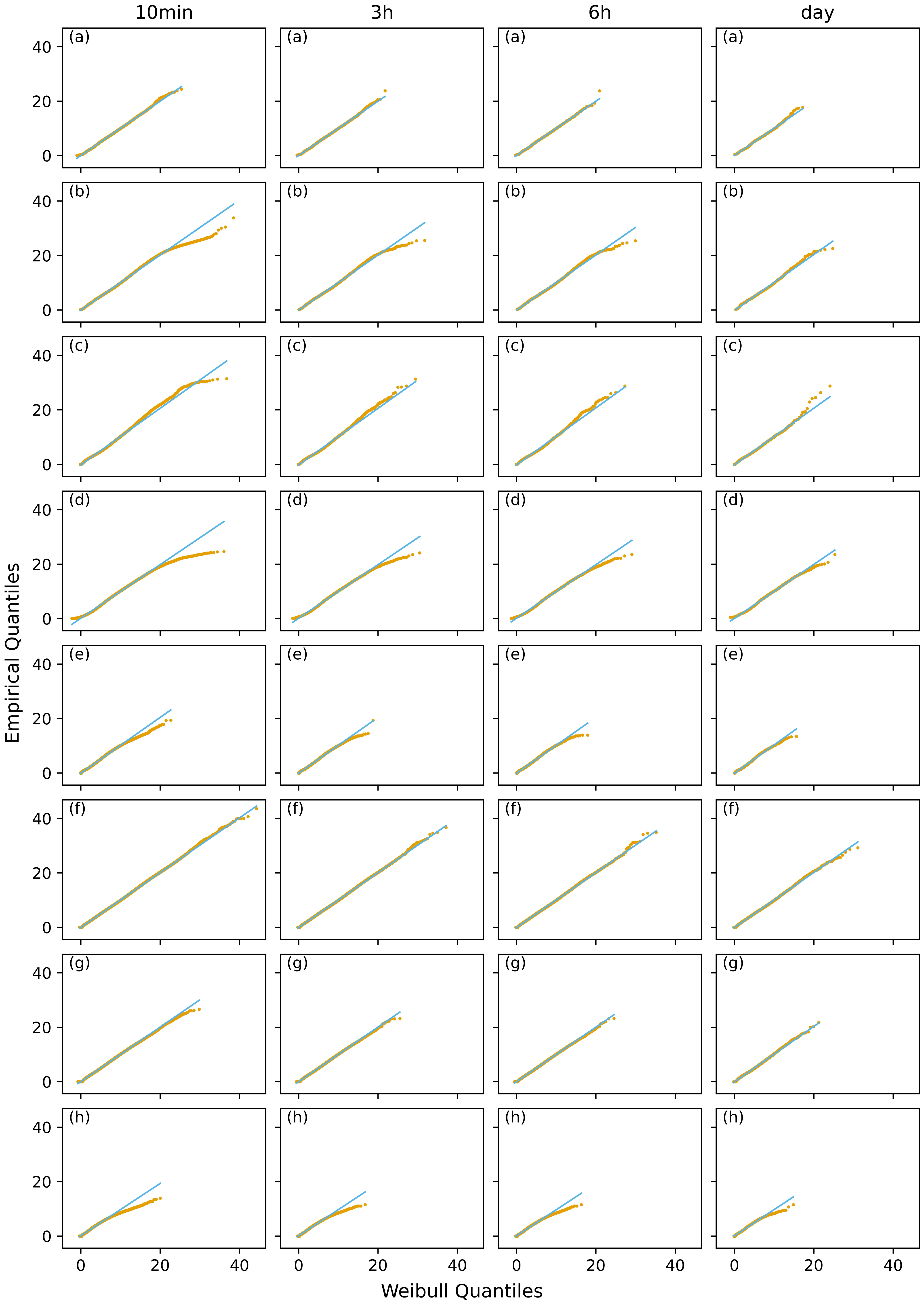}
    \caption{QQ-plots of instantaneous wind speeds and fitted Gamma distributions. From left to right the data resolution is 10min, three-hourly averages, six-hourly averages and daily averages.}
    \label{appendix:QQ-plot-Gamma-inst}
\end{figure}

\begin{figure}[H]
\begin{subfigure}{0.5\textwidth}
    \centering
    \includegraphics[width=\textwidth]{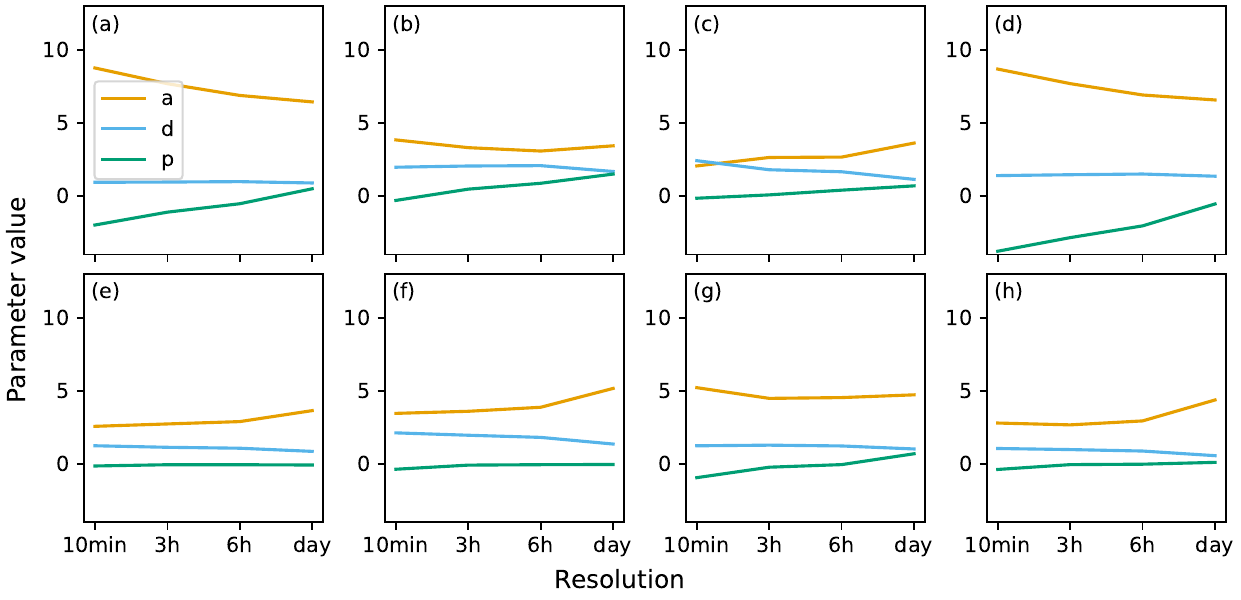}
    \caption{Absolute trends.}
\end{subfigure}
\begin{subfigure}{0.5\textwidth}
    \centering
    \includegraphics[width=\textwidth]{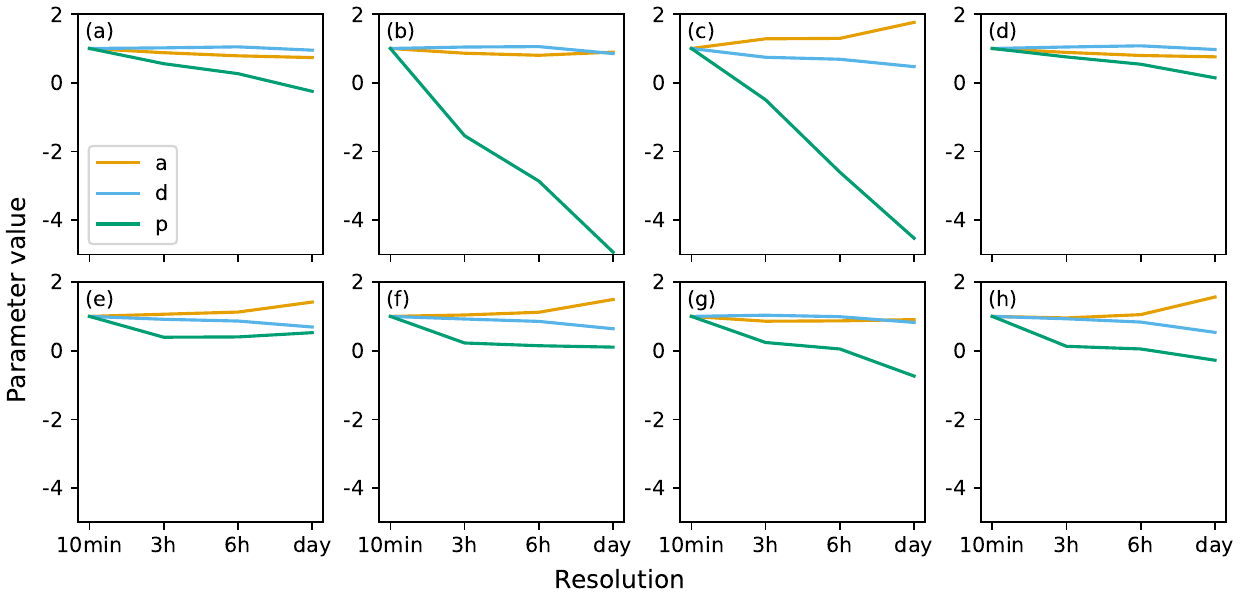}
    \caption{Relative trends. }
\end{subfigure}
    \caption{Gamma parameter trends when fitted to averaged wind speed distributions.}
        \label{appendix:gamma-rel}

\end{figure}
\begin{figure}[H]
\begin{subfigure}{.5\textwidth}
    \centering
    \includegraphics[width=\textwidth]{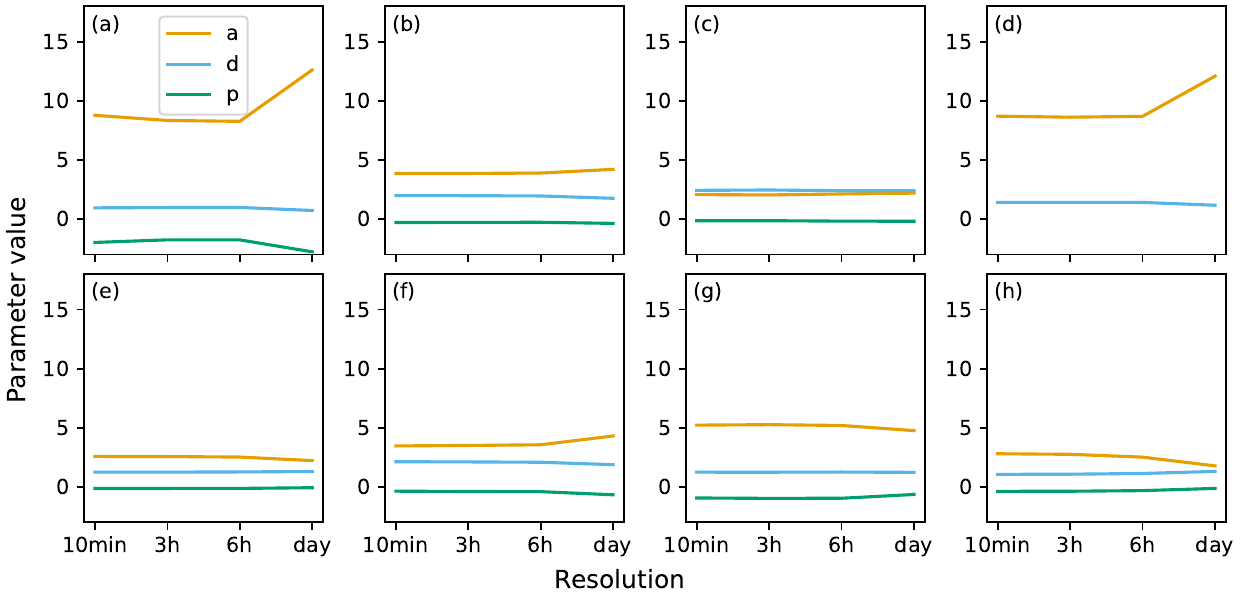}
    \caption{Absolute trends.}
\end{subfigure}
\begin{subfigure}{.5\textwidth}
    \centering
    \includegraphics[width=\textwidth]{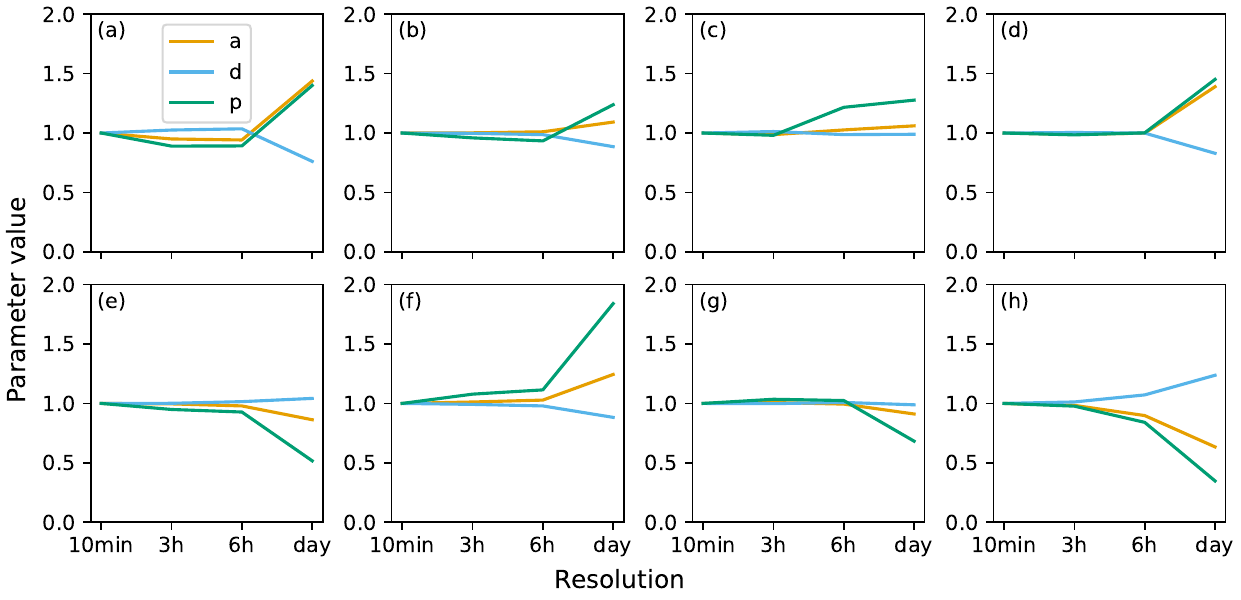}
    \caption{Relative trends.}
\end{subfigure}
\caption{Gamma parameter trends when fitted to instantaneous wind speed distributions.}
\label{appendix:gamma-inst-rel}
\end{figure}

\newpage

\begin{table}[H]
    \centering
    \small
    \begin{tabular}{|c|c|c|c|c|c|}
    \cline{2-5}
    \multicolumn{1}{c|}{}  & daily & six-hourly & three-hourly  & 10min \\
    \hline
    daily&1&$2.24\cdot10^{-3}$&$4.30\cdot10^{-5}$& $4.98\cdot10^{-9}$ \\ \hline
     six-hourly&&1&$\mathbf{4.54\cdot10^{-1}}$&$6.88\cdot10^{-6}$ \\ \hline
    three-hourly&&&1& $1.85\cdot10^{-4}$ \\ \hline
    \end{tabular}
    \caption{$p$-values of Kolmogorov-Smirnov test for averages Kelmarsh. All values where $p \leq 0.05$ are marked \textbf{bold} and the corresponding distributions are not considered to be significantly different.}
    \label{appendix:KS_kelmarsh_avrg}
\end{table}
\begin{table}[H]
    \centering
     \small
    \begin{tabular}{|c|c|c|c|c|c|}
    \cline{2-5}
    \multicolumn{1}{c|}{}  & daily & six-hourly & three-hourly & 10min \\
    \hline
    daily&1&$1.54\cdot10^{-2}$&$5.58\cdot10^{-3}$& $2.15\cdot10^{-3}$ \\ \hline
     six-hourly&&1&$\mathbf{1.00}$&$\mathbf{9.81\cdot10^{-1}}$ \\ \hline
    three-hourly&&&1& $\mathbf{1.00}$ \\ \hline
    \end{tabular}
    \caption{$p$-values of Kolmogorov-Smirnov test for instantaneous Kelmarsh. All values where $p \leq 0.05$ are marked \textbf{bold} and the corresponding distributions are not considered to be significantly different.}
    \label{appendix:KS_kelmarsh_inst}
\end{table}

\begin{table}[H]
    \centering
    \small
    \begin{tabular}{|c|c|c|c|c|c|}
    \cline{2-5}
    \multicolumn{1}{c|}{} & daily & six-hourly & three-hourly & 10min \\
    \hline
    daily&1&$8.15\cdot10^{-11}$&$9.13\cdot10^{-18}$& $1.35\cdot10^{-33}$ \\ \hline
     six-hourly&&1&$5.53\cdot10^{-4}$&$2.51\cdot10^{-35}$ \\ \hline
    three-hourly&&&1& $2.45\cdot10^{-27}$ \\ \hline
    \end{tabular}
    \caption{$p$-values of Kolmogorov-Smirnov test for averaged NWTC wind speed distributions. All values where $p \leq 0.05$ are marked \textbf{bold} and the corresponding distributions are not considered to be significantly different.}
    \label{appendix:KS_nwtc_avrg}
\end{table}

\begin{table}[H]
    \centering
    \small
    \begin{tabular}{|c|c|c|c|c|c|}
    \cline{2-5}
    \multicolumn{1}{c|}{} & daily & six-hourly & three-hourly & 10min \\
    \hline
    daily&1&$\mathbf{1.64\cdot10^{-1}}$&$3.84\cdot10^{-2}$& $3.61\cdot10^{-2}$ \\ \hline
    six-hourly&&1&$\mathbf{7.21\cdot10^{-1}}$&$\mathbf{6.20\cdot10^{-1}}$ \\ \hline
    three-hourly&&&1& $\mathbf{9.01\cdot10^{-1}}$ \\ \hline

    \end{tabular}
    \caption{$p$-values of Kolmogorov-Smirnov test for instantaneous NWTC wind speed distributions. All values where $p \leq 0.05$ are marked \textbf{bold} and the corresponding distributions are not considered to be significantly different.}
    \label{appendix:KS_nwtc_inst}
\end{table}

\begin{table}[H]
    \centering
    \small
    \begin{tabular}{|c|c|c|c|c|c|}
    \cline{2-5}
    \multicolumn{1}{c|}{} & daily & six-hourly & three-hourly &  10min \\
    \hline
    daily&1&$3.00\cdot10^{-3}$&$2.25\cdot10^{-4}$& $1.40\cdot10^{-5}$ \\ \hline
    six-hourly&&1&$\mathbf{7.15\cdot10^{-1}}$&$\mathbf{9.20\cdot10^{-2}}$ \\ \hline
    three-hourly&&&1& $\mathbf{4.95\cdot10^{-1}}$ \\ \hline
    \end{tabular}
    \caption{$p$-values of Kolmogorov-Smirnov test for averaged Owez wind speed distributions. All values where $p \leq 0.05$ are marked \textbf{bold} and the corresponding distributions are not considered to be significantly different.}
    \label{appendix:KS_owez_avrg}
\end{table}
\begin{table}[H]
    \centering
    \small
    \begin{tabular}{|c|c|c|c|c|c|}
    \cline{2-5}
    \multicolumn{1}{c|}{} & daily & six-hourly & three-hourly & 10min \\
    \hline
    daily&1&$\mathbf{2.14\cdot10^{-1}}$&$\mathbf{2.18\cdot10^{-1}}$& $\mathbf{1.29\cdot10^{-1}}$ \\ \hline
    six-hourly&&1&$\mathbf{1.00}$&$\mathbf{9.78\cdot10^{-1}}$ \\ \hline
    three-hourly&&&1& $\mathbf{9.77\cdot10^{-1}}$ \\ \hline

    \end{tabular}
    \caption{$p$-values of Kolmogorov-Smirnov test for instantaneous Owez wind speed distributions. All values where $p \leq 0.05$ are marked \textbf{bold} and the corresponding distributions are not considered to be significantly different.}
    \label{appendix:KS_owez_inst}
\end{table}

\begin{table}[H]
    \centering
    \small
    \begin{tabular}{|c|c|c|c|c|c|}
    \cline{2-5}
    \multicolumn{1}{c|}{} & daily & six-hourly & three-hourly & 10min \\
    \hline
    daily&1&$3.83\cdot10^{-14}$&$1.43\cdot10^{-21}$& $3.24\cdot10^{-52}$ \\ \hline
    six-hourly&&1&$6.11\cdot10^{-4}$&$1.76\cdot10^{-45}$ \\ \hline
    three-hourly&&&1& $1.73\cdot10^{-44}$ \\ \hline

    \end{tabular}
    \caption{$p$-values of Kolmogorov-Smirnov test for averaged Aachen wind speed distributions. All values where $p \leq 0.05$ are marked \textbf{bold} and the corresponding distributions are not considered to be significantly different.}
    \label{appendix:KS_aachen_avrg}
\end{table}
\begin{table}[H]
    \centering
    \small
    \begin{tabular}{|c|c|c|c|c|c|}
    \cline{2-5}
    \multicolumn{1}{c|}{} & daily & six-hourly & three-hourly & 10min \\
    \hline
    daily&1&$2.89\cdot10^{-22}$&$1.54\cdot10^{-25}$& $2.14\cdot10^{-28}$ \\ \hline
    six-hourly&&1&$\mathbf{9.35\cdot10^{-1}}$&$\mathbf{7.85\cdot10^{-1}}$ \\ \hline
    three-hourly&&&1& $\mathbf{1.00}$ \\ \hline

    \end{tabular}
    \caption{$p$-values of Kolmogorov-Smirnov test for instantaneous Aachen wind speed distributions. All values where $p \leq 0.05$ are marked \textbf{bold} and the corresponding distributions are not considered to be significantly different.}
    \label{appendix:KS_aachen_inst}
\end{table}

\begin{table}[H]
    \centering
    \small
    \begin{tabular}{|c|c|c|c|c|c|}
    \cline{2-5}
    \multicolumn{1}{c|}{} & daily & six-hourly & three-hourly & 10min \\
    \hline
    daily&1&$4.47\cdot10^{-23}$&$9.69\cdot10^{-39}$& $2.71\cdot10^{-72}$ \\ \hline
    six-hourly&&1&$1.16\cdot10^{-5}$&$3.01\cdot10^{-45}$ \\ \hline
    three-hourly&&&1& $4.90\cdot10^{-33}$ \\ \hline

    \end{tabular}
    \caption{$p$-values of Kolmogorov-Smirnov test for averaged Zugspitze wind speed distributions. All values where $p \leq 0.05$ are marked \textbf{bold} and the corresponding distributions are not considered to be significantly different.}
    \label{appendix:KS_zugspitze_avrg}
\end{table}
\begin{table}[H]
    \centering
    \small
    \begin{tabular}{|c|c|c|c|c|c|}
    \cline{2-5}
    \multicolumn{1}{c|}{} & daily & six-hourly & three-hourly &  10min \\
    \hline
    daily&1&$5.68\cdot10^{-22}$&$3.59\cdot10^{-24}$& $3.84\cdot10^{-27}$ \\ \hline
    six-hourly&&1&$\mathbf{7.13\cdot10^{-1}}$&$\mathbf{3.89\cdot10^{-1}}$ \\ \hline
    three-hourly&&&1& $\mathbf{1.00}$ \\ \hline

    \end{tabular}
    \caption{$p$-values of Kolmogorov-Smirnov test for instantaneous Zugspitze wind speed distributions. All values where $p \leq 0.05$ are marked \textbf{bold} and the corresponding distributions are not considered to be significantly different.}
    \label{appendix:KS_zugspitze_inst}
\end{table}

\begin{table}[H]
    \centering
    \small
    \begin{tabular}{|c|c|c|c|c|c|}
    \cline{2-5}
    \multicolumn{1}{c|}{} & daily & six-hourly & three-hourly &  10min \\
    \hline
    daily&1&$9.77\cdot10^{-34}$&$1.04\cdot10^{-64}$& $1.39\cdot10^{-85}$ \\ \hline
    six-hourly&&1&$1.04\cdot10^{-5}$&$1.69\cdot10^{-44}$ \\ \hline
    three-hourly&&&1& $4.23\cdot10^{-35}$ \\ \hline

    \end{tabular}
    \caption{$p$-values of Kolmogorov-Smirnov test for averaged Boltenhagen wind speed distributions. All values where $p \leq 0.05$ are marked \textbf{bold} and the corresponding distributions are not considered to be significantly different.}
    \label{appendix:KS_boltenhagen_avrg}
\end{table}
\begin{table}[H]
    \centering
    \small
    \begin{tabular}{|c|c|c|c|c|c|}
    \cline{2-5}
    \multicolumn{1}{c|}{} & daily & six-hourly & three-hourly &  10min \\
    \hline
    daily&1&$7.14\cdot10^{-31}$&$1.72\cdot10^{-34}$& $1.09\cdot10^{-38}$ \\ \hline
    six-hourly&&1&$\mathbf{7.84\cdot10^{-1}}$&$\mathbf{3.59\cdot10^{-1}}$ \\ \hline
    three-hourly&&&1& $\mathbf{1.00}$ \\ \hline

    \end{tabular}
    \caption{$p$-values of Kolmogorov-Smirnov test for instantaneous Boltenhagen wind speed distributions. All values where $p \leq 0.05$ are marked \textbf{bold} and the corresponding distributions are not considered to be significantly different.}
    \label{appendix:KS_boltenhagen_inst}
\end{table}

\begin{table}[H]
    \centering
    \small
    \begin{tabular}{|c|c|c|c|c|c|}
    \cline{2-5}
    \multicolumn{1}{c|}{} & daily & six-hourly & three-hourly &  10min \\
    \hline
    daily&1&$4.20\cdot10^{-108}$&$2.04\cdot10^{-163}$& $3.60\cdot10^{-265}$ \\ \hline
    six-hourly&&1&$1.47\cdot10^{-17}$&$1.64\cdot10^{-187}$ \\ \hline
    three-hourly&&&1& $6.61\cdot10^{-184}$ \\ \hline

    \end{tabular}
    \caption{$p$-values of Kolmogorov-Smirnov test for averaged Fichtelberg wind speed distributions. All values where $p \leq 0.05$ are marked \textbf{bold} and the corresponding distributions are not considered to be significantly different.}
    \label{appendix:KS_fichtelberg_avrg}
\end{table}
\begin{table}[H]
    \centering
    \small
    \begin{tabular}{|c|c|c|c|c|c|}
    \cline{2-5}
    \multicolumn{1}{c|}{} & daily & six-hourly & three-hourly  & 10min \\
    \hline
    daily&1&$4.34\cdot10^{-83}$&$2.06\cdot10^{-117}$& $1.27\cdot10^{-135}$ \\ \hline
    six-hourly&&1&$9.02\cdot10^{-6}$&$7.67\cdot10^{-12}$ \\ \hline
    three-hourly&&&1& $\mathbf{4.24\cdot10^{-1}}$ \\ \hline

    \end{tabular}
    \caption{$p$-values of Kolmogorov-Smirnov test for instantaneous Fichtelberg wind speed distributions. All values where $p \leq 0.05$ are marked \textbf{bold} and the corresponding distributions are not considered to be significantly different.}
    \label{appendix:KS_fichtelberg_inst}
\end{table}

\begin{table}[H]
    \centering
    \small
    \begin{tabular}{|l|c|c|c|c|}\hline
        Location & 10min & 3h avrg. & 3h inst. & 6h inst. \\ \hline
         Aachen& 0 & -40.28 & -3.47 &1.99 \\\hline
         Zugspitze& 0 & -39.16& 4.76 & 6.06 \\\hline
         Boltenhagen &0 & -111.23& 9.86 & 3.69 \\\hline
         Fichtelberg & 0 & -62.71 & 1.83& -9.80 \\\hline
         
    \end{tabular}
    \caption{Errors (in $MW$) of wind power generation prediction when using 30 years of data of lower resolution (three-hourly average and three-hourly and six-hourly instantaneous) wind speed observations compared to 10min wind speed observations. }
    \label{appendix:errors}
\end{table}

\begin{comment}
  \begin{table}[H]
    \centering
    \begin{tabular}{l|c|c}
    \hline
         Station& Lon, Lat & Closest CMIP6 Lon, Lat \\ \hline
         Kelmarsh & 52.40, -0.95 & 53.16, 358.10 \\ \hline
         Penmanshiel & 55.90, -2.31 & 55.02, 358.10 \\ \hline
         Owez & 52.61, 4.39 & 53.16, 3.75\\ \hline
         NWTC & 39.21, -105.23 & 40.10, 255.00\\ \hline
         Aachen & 50.78, 6.09 &  51.29, 5.63 \\ \hline
         Boltenhagen & 54.00, 11.19 & 53.16, 11.25 \\ \hline
         Fichtelberg & 49.98, 11.84 & 49.43, 11.25 \\ \hline
         Zugspitze & 47.42, 10.98 & 47.56, 11.25 \\ \hline
    \end{tabular}
    \caption{\textcolor{red}{Do I need this information? Looks weird. Currently not referenced.}}
    \label{appendix:closest-CMIP}
\end{table}  
\end{comment}